\documentclass{aa}  

\usepackage[toc,page]{appendix}
\usepackage{graphicx}
\usepackage{txfonts}
\usepackage{natbib}
\usepackage[caption=false]{subfig}
\usepackage{float}
\usepackage[colorlinks=true,urlcolor=blue,citecolor=blue,linkcolor=blue]{hyperref}
\usepackage{xcolor}

\newcommand{\kms}{\mbox{km~s$^{-1}$}}

\begin{document} 

\title{The formation and heating of chromospheric fibrils in a radiation-MHD simulation}


   \author{M. K. Druett\inst{1}
          \and
          J. Leenaarts\inst{1}
          \and
          M. Carlsson\inst{2,3}
          \and
          M. Szydlarski\inst{2,3}}

\offprints{M. K. Druett \email{malcolm.druett@astro.su.se}}

   \institute{Institute for Solar Physics, Department of Astronomy, Stockholm University, AlbaNova University Centre, SE-106 91 Stockholm, Sweden
   \and
Institute of Theoretical Astrophysics, University of Oslo, P.O. Box 1029 Blindern, N-0315 Oslo, Norway
\and
Rosseland Centre for Solar Physics, University of Oslo, P.O. Box 1029 Blindern, N-0315 Oslo, Norway}

    \date{\today}

 
  \abstract
   {}
   {We examine the movements of mass elements within dense fibrils using passive tracer particles (corks) in order to understand fibril creation and destruction processes.}
   {Simulated fibrils were selected at times when they were visible in an H$\alpha$ image proxy. The corks were selected within fibril H$\alpha$ formation regions. From this set, a cork was selected, and the field line passing through it was constructed. Other fibrilar corks close to this fieldline were also selected. Pathlines were constructed, revealing the locations of the mass elements forward and backward in time. The forces acting on these mass elements were analysed.}
   {The main process of fibrilar loading in the simulation is different to the mass loading scenario in which waves steepen into shocks and push material upwards along the fieldlines from near their footpoints. Twisted low lying fieldlines were destabilised and then they untwisted, lifting the material trapped above their apexes via the Lorentz force. Subsequently, the majority of the mass drained down the fieldlines towards one or both footpoints under gravity. Material with large horizontal velocities could also elevated in rising fieldlines, creating somewhat parabolic motions, but material was not generally moving upward along a stationary magnetic fieldline during loading.}
   {The processes observed in the simulation are plausible additional scenarios. Criteria for observing such events are described. It is desirable that our simulations can also form more densely-packed fibrils from material fed from the base of field footpoints. Experimental parameters required to achieve this are discussed.}
   \keywords{ Sun: chromosphere -- Sun: magnetic fields -- Sun: transition region -- Magnetohydrodynamics (MHD) -- Plasmas}

   \maketitle

   \section{Introduction} \label{sec:Intr}



Fibrils are long ($500-20000$~km), narrow ($\sim180$~km) fibre-like features, with lifetimes of around $200-400$~s, that are seen in chromospheric lines \citep{2007DePontieu2Spicules, 2017GafeiraMorphologyFibrils, 2017JafarzadehFibrilMapping}. 
Fibrils are ubiquitously observed on the Sun - in magnetic bright points, plages, sunspots, and regions of the quiet Sun that have any significant photospheric concentrations of magnetic field, which can act as the roots of the structures \citep{2010WiegelmannLoops}. They are generally narrower in Active regions, and broader in the quiet Sun. They can be bright or dark in appearance, and have been observed in H$\alpha$ \citep{2006Hansteen, 2015LeenaartsFibrils, 2017MooroogenFibrils, 2018GopalanPriyaFibrils}, Ca II 8542 \citep{2017AsensioRamosFibrilsCa8542}, Ca II H \& K \citep{2017GafeiraMorphologyFibrils, 2017JafarzadehFibrilMapping, 2020Kianfar}, He I 10830 \citep{2013SchadFibrilsHeI10830}, He I D$_3$ triplet \citep{2019LibbrechtHeID3}, He II 304 \citep{2014ZhangFibrilFeedingProminenceHeII}, and Ly$\alpha$ \citep{2017RuttenFibrilsLyAlphaALMA} to name a few.


Dark H$\alpha$ fibrils are thought to be the observational signatures of locations with high ridges of increased mass density at chromospheric temperatures. Under these conditions the H$\alpha$ line optical depth reaches unity at greater physical heights above the photosphere because the H$\alpha$ line opacity is relatively insensitive to typical chromospheric temperature variations, but depends heavily on mass density. 

The combination of this opacity variation with source function decrease with height and scattering dominated 3D radiative transfer effects produces a negative correlation of emission intensity in H$\alpha$ with increased line core formation height \cite{2015LeenaartsFibrils}.

Because the density of the material in fibrils is so much greater than that of the surrounding material, fibrils are essential to understanding the balance of mass throughout the solar atmosphere. Thus it is important to get a clear picture of the sources of this mass and the forces loading it into these fibrilar structures.

Fibrils also play an important role in the transport of energy from the photosphere to the chromosphere and possibly also the corona \citep{2017GafeiraMorphologyFibrils} via wave propagation \citep{2007CarlssonWaveHeating, 2014MortonFibril} or small scale reconnection \citep{1972Parker}. 


Several scenarios to explain mass loading of fibrils exist, based on observational evidence and earlier models. The standard mass loading scenario is that p-mode oscillations in the photosphere drive compressive longitudinal waves that are guided by the magnetic field and feed mass up relatively static fieldlines \citep{2006Hansteen, 2007DePontieu2Spicules}. 
More recently \cite{2019Rutten} suggested that some dark H$\alpha$ fibrils are the cooling aftermaths of heating events associated with upward jets.
They are also thought to form around the boundaries of active regions due to the rising horizontal field seen in areas of late stage flux emergence \citep{2002Bernasconi,2002Mandrini, 2010Xu}. Some apparent motions of material in chromospheric fine structure are not so well explained by motions confined to narrow tubes of plasma. \cite{2011Judge} show that these can be better attributed to opacity effects due to the line of sight ripples in what are essentially 2D sheets structures with magnetic tangential discontinuities, like the folds in a semi-transparent curtain moving in a breeze. In particular this theory can explain observational reports \citep{2012Judge} of extremely high apparent velocities of features that greatly exceed local Alfv\'{e}n speeds.

Therefore, in this paper we address several unanswered questions regarding mass loading of fibrils using simulations:
What source(s) supply fibrils with their mass?
What are the physical mechanisms/forces that load this mass into the fibrils?
What are the dominant forces and energy sources acting on the plasma elements in fibrils, and what changes do these bring about?
What processes can destroy fibrils?
To which locations does the mass that is contained in a fibril flow?


\section{Experiment description} \label{sec:ExpeDesc}
  \begin{figure*} 
   \includegraphics[width=18cm]{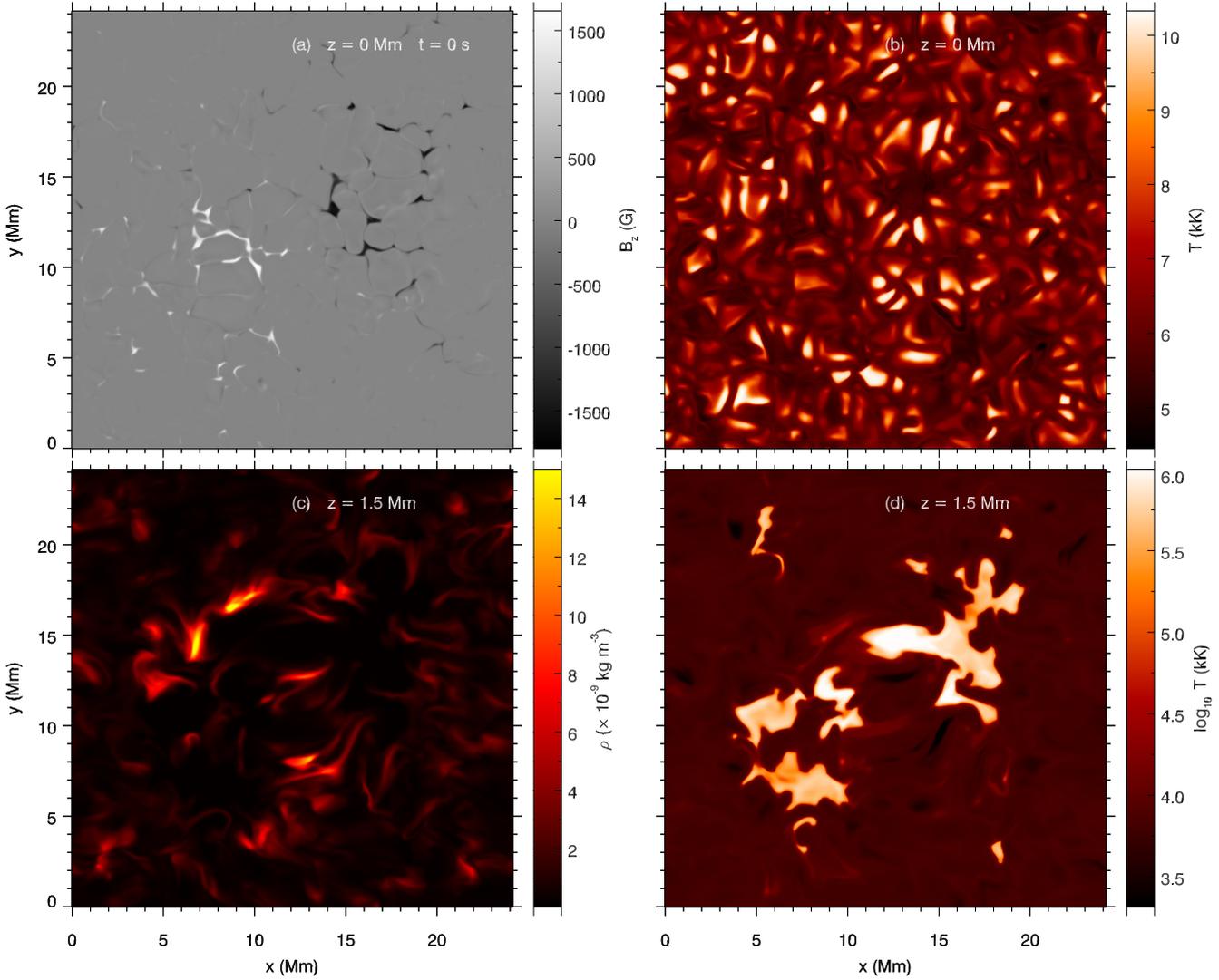}
   \caption{Simulation setup, with vertical magnetic field (a) and temperature (b) in the photosphere at $z=0$~Mm, and mass density (c) and temperature (d) at $z=1.5$~Mm in the chromosphere.}
   \label{fig:experiment}
   \end{figure*}

\subsection{Bifrost experiment description} \label{sec:bifr}

The simulation for this work was performed in 3D using the Bifrost code \citep{2011GudiksenBifrost}, which solves the resistive MHD equations on a staggered Cartesian mesh. Additional modules in the simulation included optically thick radiative transfer in the photosphere and low chromosphere, parameterised radiative losses in the upper chromosphere, transition region and corona, and thermal conduction along magnetic field lines. 

\begin{figure}
   \centering
   \includegraphics[width=8.8cm]{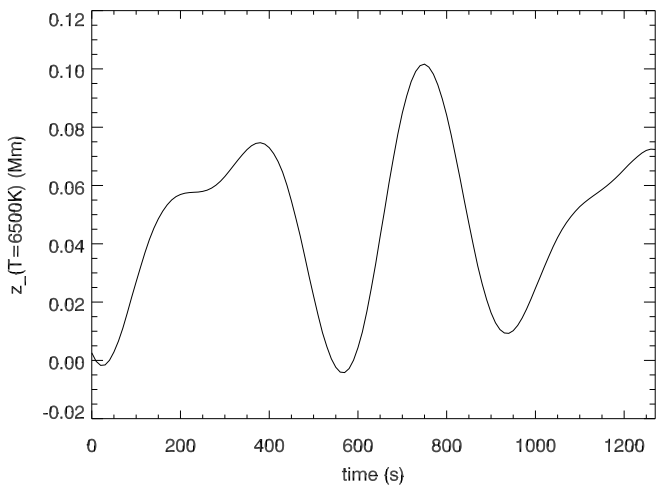}
   \caption{Box oscillations in the simulation. The black curve shows the time evolution of the height where the average temperature is 6500~K in the photosphere. These oscillations are the simulation equivalent to the p-mode oscillations of the photosphere on the Sun.}
   \label{fig:box_osc}
\end{figure}

The simulation was run on a re-sampled mesh based on the one used in \cite{2016Carlsson} and for two studies of fibrils in \cite{2012LeenaartsFibrils, 2015LeenaartsFibrils}, namely $504 \times 504 \times 496$ grid cells. In our experiment the vertical number of grid cells increased to 512 in order to help improve code stability. The horizontal extent of the model is of $24 \times 24 \times 16.8$~Mm. The vertical grid spans from 2.4~Mm below to 14.4~Mm above average optical depth unity at 500~nm, thereby encompassing the upper convection zone, photosphere, chromosphere, and lower corona. The $x$- and $y$-axes are equidistant with a grid spacing of 48~km. The $z$-axis grid spacing is 20~km between -1 and 4.5~Mm, but this spacing increases toward the upper and lower boundaries reaching a maximum of 98~km, in the corona. Fig\ref{fig:experiment} shows a snapshot of the experimental setup. As in \cite{2016Carlsson, 2012LeenaartsFibrils, 2015LeenaartsFibrils, 2018Zacharias}, the magnetic field in the photosphere is a predominantly bipolar structure seen as two clusters of magnetic concentrations of opposite polarity. This field was introduced into the simulation by specifying the vertical magnetic field at the bottom boundary with an averaged signed field of zero and two patches of opposite polarity separated by 8~Mm. A potential field extrapolation then provided the magnetic field throughout the box. This simulation ran for 3000~s of solar time before the grid re-sampling and then a further 2400~s before the beginning of the experiment presented here. Fig.~\ref{fig:box_osc} shows the height of the experiment photosphere plotted against time in order to illustrate the timing and amplitude (around 100~km, peak-to-peak) of the global mode of the box oscillations throughout the experiment.
\citep{2001ApJ...546..585S,2016Carlsson}.
%

A test run was conducted that also included the effects of non-equilibrium ionisation of hydrogen as in \cite{2012LeenaartsFibrils, 2015LeenaartsFibrils}. Because of the high computational expense of this physics and the fact that it was not essential for our analysis of the MHD variables in these high-density, low-temperature fibrilar structures, the module was not included for the production run.


We use the Lagrangian tracer particles (hereafter referred to simply as ''corks'') module described in \cite{2018LeenaartsCorks} to analyse the simulation. Through regular injection of new corks in cork-free voids and removal of corks in regions where they pile up, this module allows tracing of velocity pathlines starting from any time and at any location in the simulation.  A short test experiment confirmed that the accuracy of both backward and forward traced pathlines in fibrils is accurate to within a grid-cell, see appendix \ref{sec:appcorks} for details. 

The cork positions are written to snapshot files at 10~s intervals. The cork motions are determined from instantaneous values at each hydrodynamic timestep. Instantaneous variable values for the corks used in our analysis are re-calculated from the values saved for the experiment grid at each snapshot. These variables include, the densities, magnetic flux densities (hereafter referred to as the magnetic field strength), pressures and temperatures, momenta, $x$-$y$-$z$ force components (from the Lorentz force, numerical momentum and mass diffusion, pressure gradient, and viscous stress forces) as well as heating quantities (from energy diffusion, Joule heating, viscous heating, radiation, and Spitzer conduction). This implies that influences acting in shorter timescales than the snapshots may not be accurately captured in our analysis, although they will affect the motions of the corks. A shorter experiment was conducted with snapshots saved every 1~s. It was found that occasionally there were relevant forces acting on shorter timescales, in particular the action of the Lorentz force, but that the general dynamics of the structures we are studying can be inferred from the data saved at 10~s intervals. 

\subsection{Selection of the corks contained in fibrils} \label{sec:selectcork}

\begin{figure*}
   \includegraphics[width=18cm]{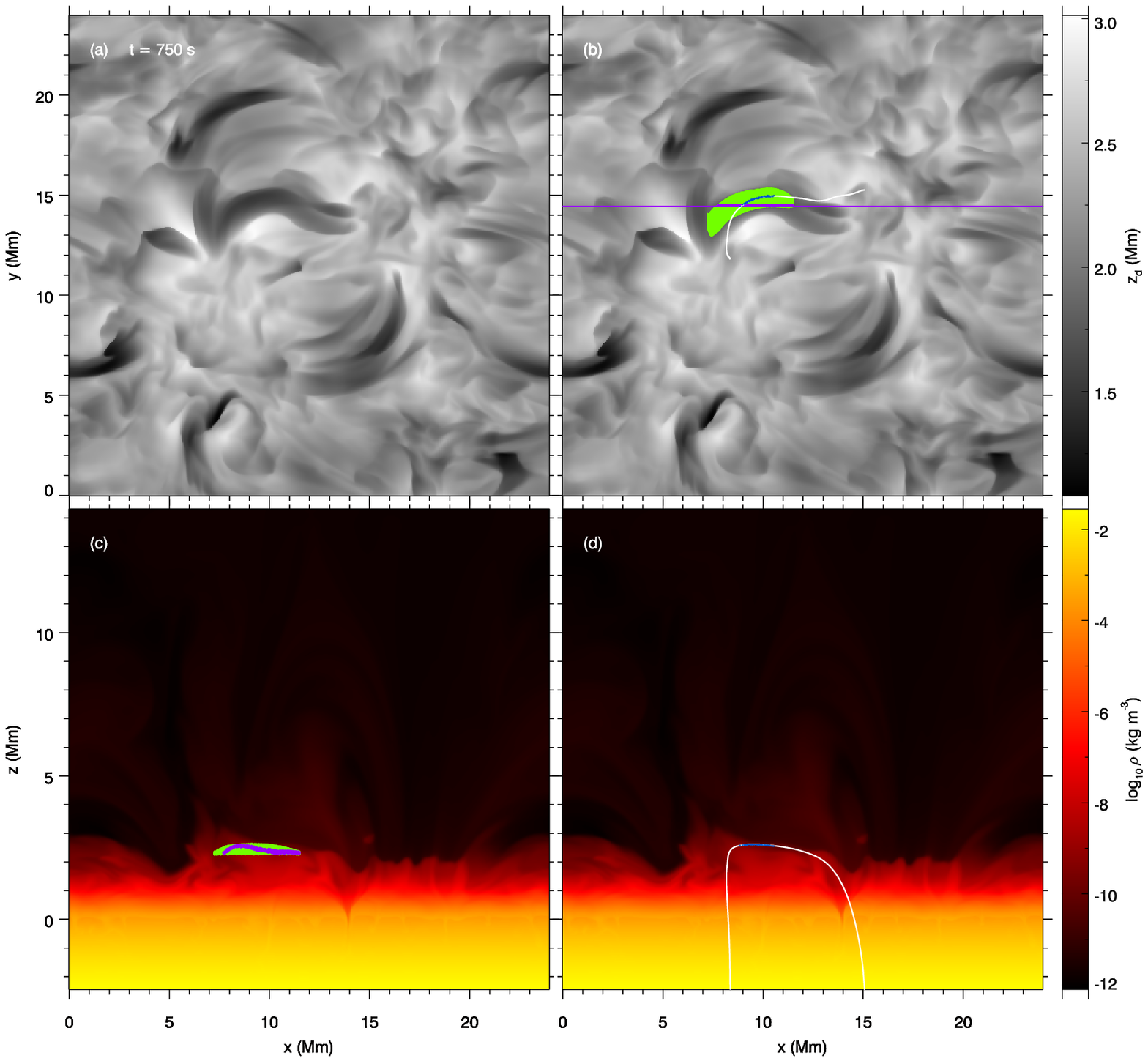}
   \caption{Example of selection of corks in a fibril  at $t=750$~s. The H$\alpha$ line intensity proxy is shown in panels a and b. Panels c and d  display the density in the $xz$-plane  along the magenta line in panel b. The locations of all corks in a fibril are indicated in green, while the fibrilar corks located along the $xz$-cut are indicated in magenta in panel c. The white curve shows a magnetic field line passing through the fibril; fibrilar corks located on or very close to the field line are shown in blue.}
   \label{fig:select}
\end{figure*}

Fibrils are fundamentally an observed phenomenon in a number of chromospheric lines. In order to build on results from \cite{2012LeenaartsFibrils, 2015LeenaartsFibrils} we considered fibrils that appear in the H$\alpha$ line. Due to limits on the available computing resources, this simulation was not run with non-equilibrium Hydrogen ionisation and we did not compute synthetic H$\alpha$ images. Instead we create proxy H$\alpha$ images based on the anti-correlation between H$\alpha$ line core intensity and $\tau=1$ height. The column mass at which the H$\alpha$ line core reaches optical depth unity is taken to be $3 \times 10^{-5}$~g cm$^{-2}$ as found in Fig.12 of \cite{2012LeenaartsFibrils} and the text of \cite{2015LeenaartsFibrils}.

We then proceeded to select corks in the fibrils:
For each $xy$-pixel in our simulation we computed the height for which the column mass is $3 \times 10^{-5}$~g cm$^{-2}$. The $\tau=1$ height is not always representative for the actual formation height of the line core \citep{2012LeenaartsFibrils}, and the chromosphere might not be overly dense at this height. 
Therefore, we inspect all grid cells above this height that have a temperature below 10~kK, and select the height with maximum density $z_\mathrm{d}$ as proxy for the H$\alpha$ formation height.

To select the corks in a fibril we first visually identified fibrils via the H$\alpha$ intensity proxy. We then selected $xy$-pixels in the fibril using a local intensity threshold. All corks in a given column in the fibril whose height was less than 100 km away from $z_\mathrm{d}(x,y)$ were considered to be located in the fibril.

An example of the results of this selection process is shown in Fig.~\ref{fig:select}. To demonstrate that fibrilar corks are "draped over" fibrilar density ridges, rather than occupying a solid structure, we also show corks located close to the vertical cut indicated in panel b. The green "ridge" structure shows a height range of the fibril, which is much larger than the typical span of the line core formation region of only a few hundred kilometers \citep{2015LeenaartsFibrils}.

In order to analyse forces on individual corks a further subset was selected by constructing the fieldline through a cork that was centrally located in the fibril (see panel~\ref{fig:select}d). All corks within one grid-cell of the fieldline were selected, and are shown in blue in panels (b) and (d).  Finally we constructed pathlines from $t=0$~s to $t=1270$~s for each cork in the fibril. This allows us to trace the evolution of the mass inside the fibril both forwards and backwards in time.

This procedure was repeated for twenty fibrils. Representative examples of the behaviour of three fibrils and some corks within them are presented in the next section.

\section{Results} \label{sec:results}

   \begin{figure*}
   \centering
   \includegraphics[width=\textwidth]{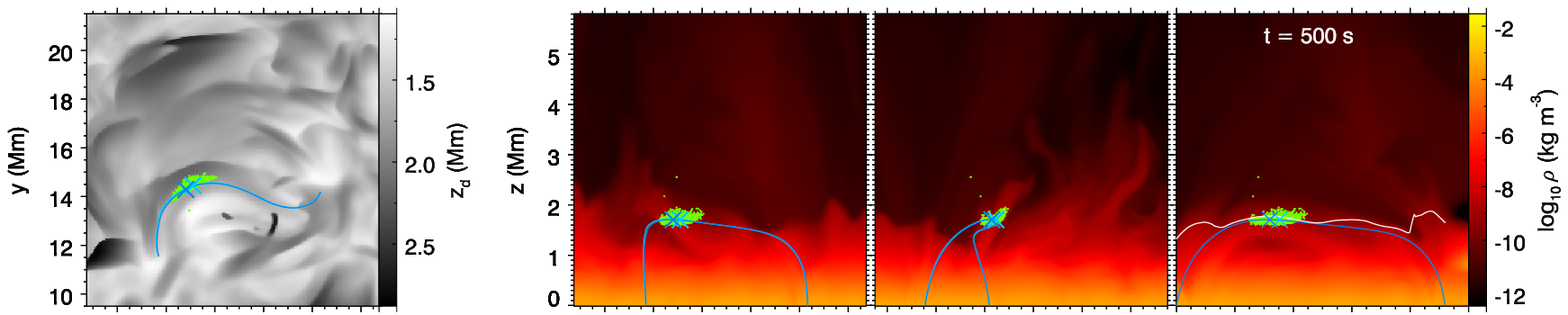}   
   \includegraphics[width=\textwidth]{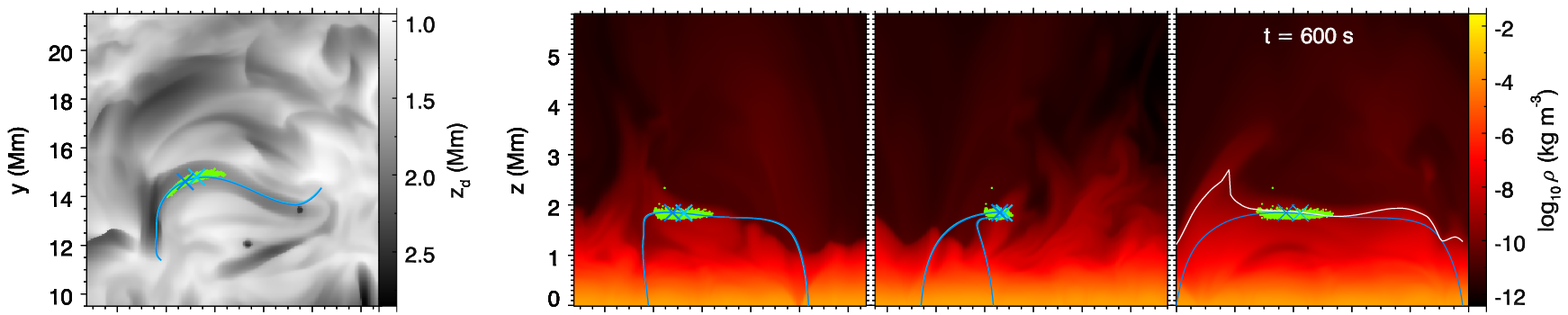}
   \includegraphics[width=\textwidth]{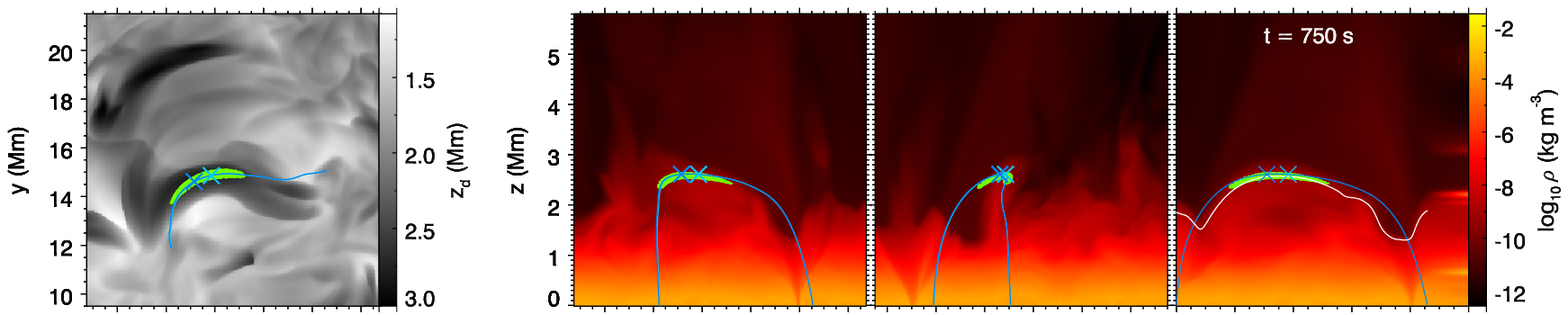} 
   \includegraphics[width=\textwidth]{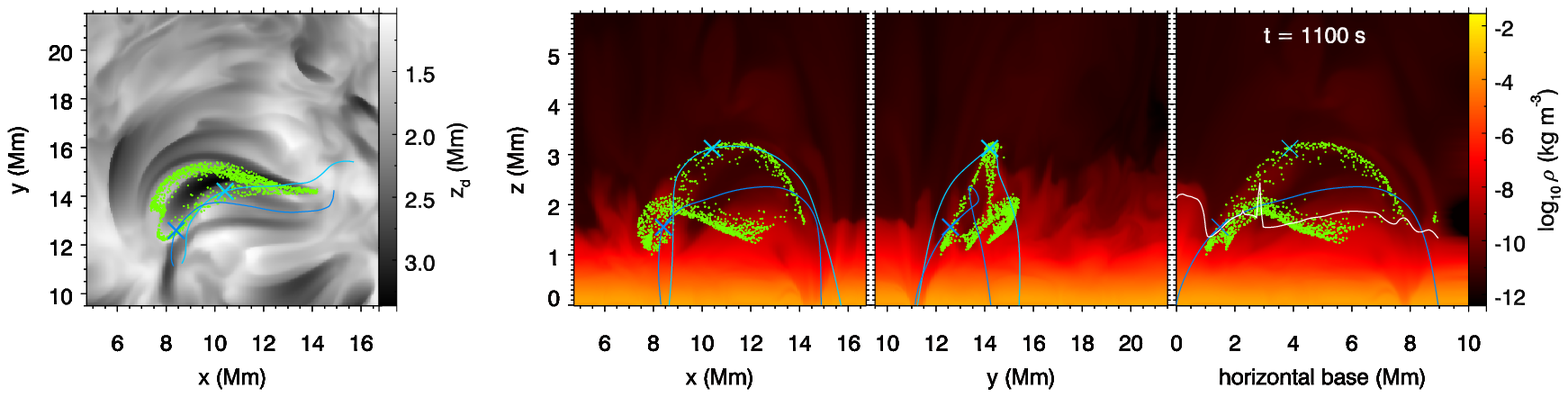}
   \caption{Time evolution of Fibril 1. Rows show from top to bottom the time evolution of the fibril, with the time indicated in the rightmost panels. The first column shows an image of the H$\alpha$ density proxy, the locations of the corks that make up the fibril at $t=750$~s, and two magnetic field lines. The field lines are initially close together, but have drifted apart at $t=1100$~s as seen in the bottom row. The second column shows the density in a vertical cut in the $xz$-plane, with the field lines and cork locations overplotted. The third column. shows the same as the second column, but for a $yz$-cut. The fourth column shows a similar vertical cut, but now traced along the $xy$ -position of one of the field lines, with the representative H$\alpha$ core formation height "z\_d" used for the H$\alpha$ proxy images overplotted using a white line.}
     \label{fig:fib1_spatial}
   \end{figure*}

\subsection{Fibril 1} \label{sec:fib1}

The first fibril that we present is a particularly thick, conspicuous, curved fibril near the centre of the field of view between the opposite polarity regions (Figs.~\ref{fig:select} and \ref{fig:fib1_spatial}). This fibril displays a formation and mass loading scenario that is common in this simulation, hereafter referred to as "lift and drain".

The fibril was very prominent at $t=750$~s, so the corks were seeded into the fibril at this time as shown in Fig.~\ref{fig:select}. 

Fig.~\ref{fig:fib1_spatial} shows the positions of the 2000 corks in the fibril closest to the fieldline at the seed time (green). Two fieldlines that pass through our corks that we selected for presentation in this paper are shown in dark blue (fibril 1, cork A) and light blue (cork B), with the locations of the corks marked with a cross.

The top row shows the locations of the corks at $t=500$~s, before the fibril formed. The corks were not located at the loop footpoints, but instead located more centrally along the fieldlines, and already on a ridge of plasma above a region of lower density (see Fig.~\ref{fig:fib1_spatial}, top right panel). Between $t=500$~s and $t=750$~s the fieldlines rose and relaxed to become less twisted, at the same time as the global box oscillation moved upwards. Beforehand the fieldline apex had flattened and twisted (Fig.~\ref{fig:fib1_spatial}, compare first and second rows), but then the central part of the fieldlines rose up to form a more semi-circular shape (Fig.~\ref{fig:fib1_spatial}, compare second and third rows). By $t=750$~s the corks had migrated towards the apex of the fibril fieldline (hence their selection at the seed time).

As the fieldlines continued to rise, a fraction of the corks start to drain towards the footpoints of the fieldlines. Another fraction of the corks actually rises, and, as we shall show below, reaches transition region temperatures. 
 
The $t=1100$~s row in Fig.~\ref{fig:fib1_spatial} shows the fates of the material after the destruction of the fibril. From the H$\alpha$ proxy panels on the left, one can see that there is still a fibril in the vicinity of the one we study, but this fibril is made up from different material to the fibril at $t=750$~s, as indicated by the mismatch between the cork positions and the fibril at $t=1100$~s.

We note that at $t=750$~s, the field lines show only a general, but not a one-to-one correspondence with the shape of the fibril, and become rapidly disassociated from its position after the fibril is destroyed.

In summary, for this "lift and drain" scenario, material along the central length of low-lying fieldlines is raised up to form high density ridges and then subsequently drains into the footpoints, rather than being fed upwards from the footpoints of relatively static fieldlines. Some material does not drain, but instead moves up into the transition region.

In Sect.~\ref{sec:F1A} we describe the evolution of a cork in Fibril~1 (which we label F1A) that ultimately drains towards a fieldline footpoint, and in Sect.~\ref{sec:F1B} we describe the evolution of cork F1B that moves into the transition region.

\subsubsection{Analysing the forces acting on a cork}

The fibrils are located in a low plasma-$\beta$ regime, where the forces and dynamics behave differently parallel and orthogonal to the magnetic field.  Therefore, we use the Frenet-Serret \citep{Serret, Frenet} coordinate frame to analyse the motions of corks with respect to the fieldlines. This frame is defined at each point along a fieldline by components in the directions of three orthogonal unit vectors as follows: $\vec{T}$ is a unit vector pointing in the direction of the magnetic field; $\vec{N}$ is a normal vector to $\vec{T}$ and points towards the centre of curvature of the fieldline; $\vec{P}$ the binormal component in the direction defined by $\vec{P} = \vec{T} \times \vec{N}$. This forms a natural set of axes to investigate motions with respect to the fieldline as longitudinal components are determined via the $\vec{T}$ component and the transverse motions are shown via the $\vec{N}$ and $\vec{P}$ components.

In  Figs.~\ref{fig:fib1a_acc}, \ref{fig:fib1b_acc}, \ref{fig:fib2_acc}, and \ref{fig:fib2b_acc}, we show the position, velocity, acceleration, and forces acting on the corks. The global box oscillation from Fig.~\ref{fig:box_osc} is shown for comparison.

The third row of panels shows the instantaneous acceleration in a given direction calculated from the net force that was saved for each snapshot (labelled {\it a\_forces}).
We also show the acceleration computed from the velocity saved in the snapshot ({\it a\_vel}). This represents an "average" acceleration over the snapshot interval time of 10~s. Finally, we show the instantaneous acceleration caused by the sum of the pressure gradient, Lorentz, and gravitational forces ({\it a\_PLG}), which are typically the dominant forces that determine the motions of the corks. In the bottom panels we show each of these three forces individually.

We note that the total acceleration from the force components shows spiked profiles in for example Fig~\ref{fig:fib1a_acc} owing to the short-term variations of the individual force components. We do not resolve this variation using the 10~s time interval with which we save the simulation state
\citep{2018LeenaartsCorks}. 
However, there are also many corks where this short term variation is small (an example is shown in Fig~\ref{fig:fib2_acc}).

The viscous stress and mass and momentum diffusion also cause a force (see Sec.~\ref{sec:bifr}). We do not show these three additional forces because they clutter rather than clarify the motions within the fibril, but their contributions can be seen from the difference between a\_Forces and a\_PLG in the third row. 

It turns out that the diffusion forces were negligible in almost all cases, but the stress force shows sharply spiked, short term variations. These spikes seem to be primarily a response to the spikes in the Lorentz force, which the viscosity acts to dampen, as can be seen from the more gentle variations of a\_forces compared to a\_PLG and its Lorentz force component, particularly during the destruction of the fibril (Fig.~\ref{fig:fib1a_acc}, around $t=1000$~s).

\subsubsection{Fibril 1, Cork A: Draining} \label{sec:F1A}

   \begin{figure*}
   \centering
   \includegraphics[width=\textwidth]{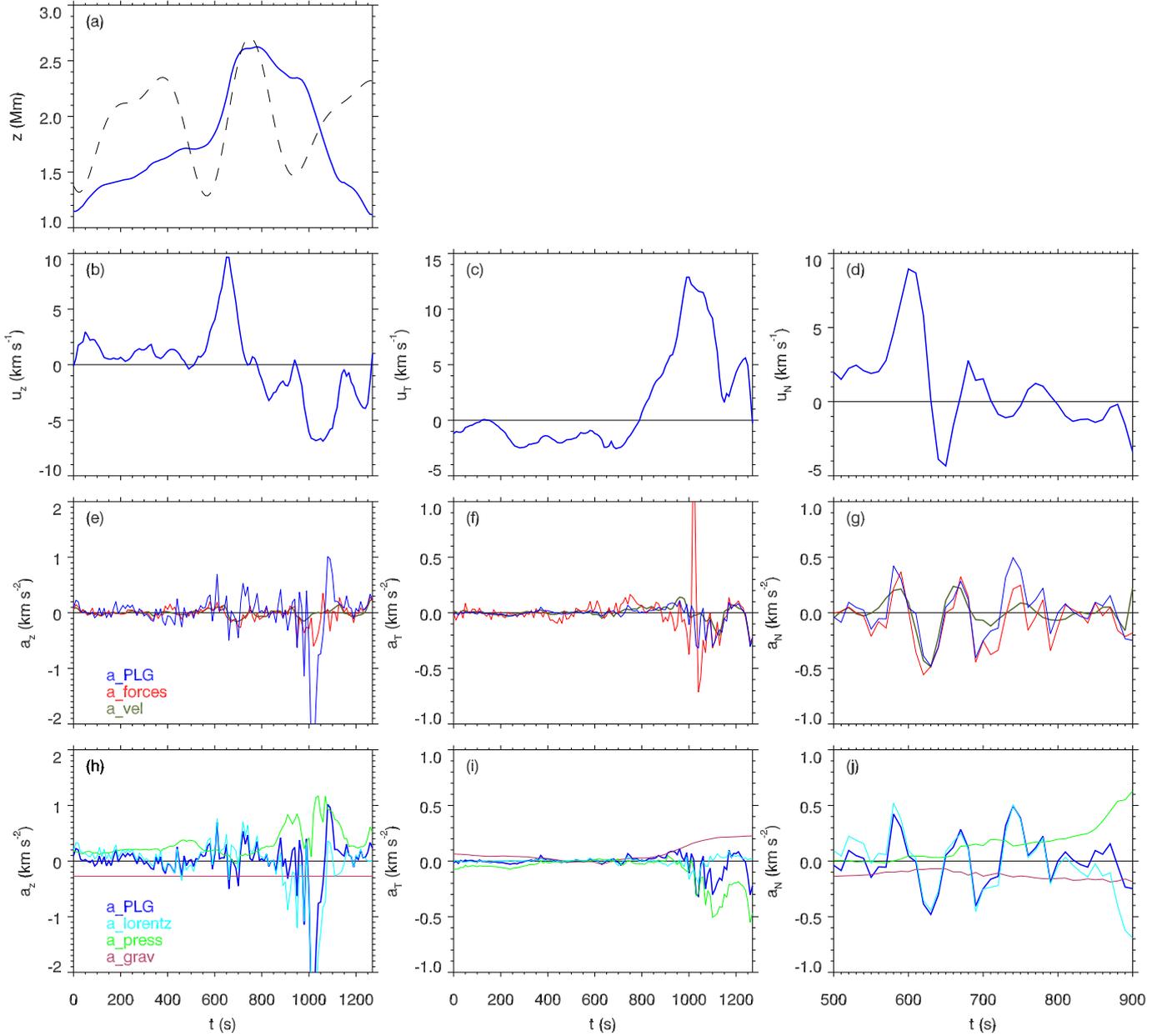}
   \caption{Positions, velocities, and accelerations for representative cork F1A in Fibril 1 as functions of time.
    {\it Top row:} cork position (solid), box oscillation for comparison (dashed, on an arbitrary scale). 
   {\it Second row:} cork velocity with a $v_z=0$ line over-plotted for reference.
    {\it Third row:} cork acceleration computed for three different assumptions. Red: computed from all forces acting on the cork; blue: computed only from the Lorentz force, gas pressure gradient, and gravity; green: computed as the time derivative of the velocity in Panel~b. 
    {\it Fourth row:} acceleration owing to the gas pressure gradient, the Lorentz force, and gravity, as well as their sum ($a_\mathrm{PLG}$). 
    {\it Left column:} in the vertical direction
    {\it Middle column:} in the direction tangential to the field line (the $\vec{T}$-direction in the Frenet-Serret frame).
    {\it Right column:} in the direction normal to the field line (the $\vec{N}$-direction).}
   \label{fig:fib1a_acc}
   \end{figure*}

   \begin{figure*}
   \centering
   \includegraphics[width=\textwidth]{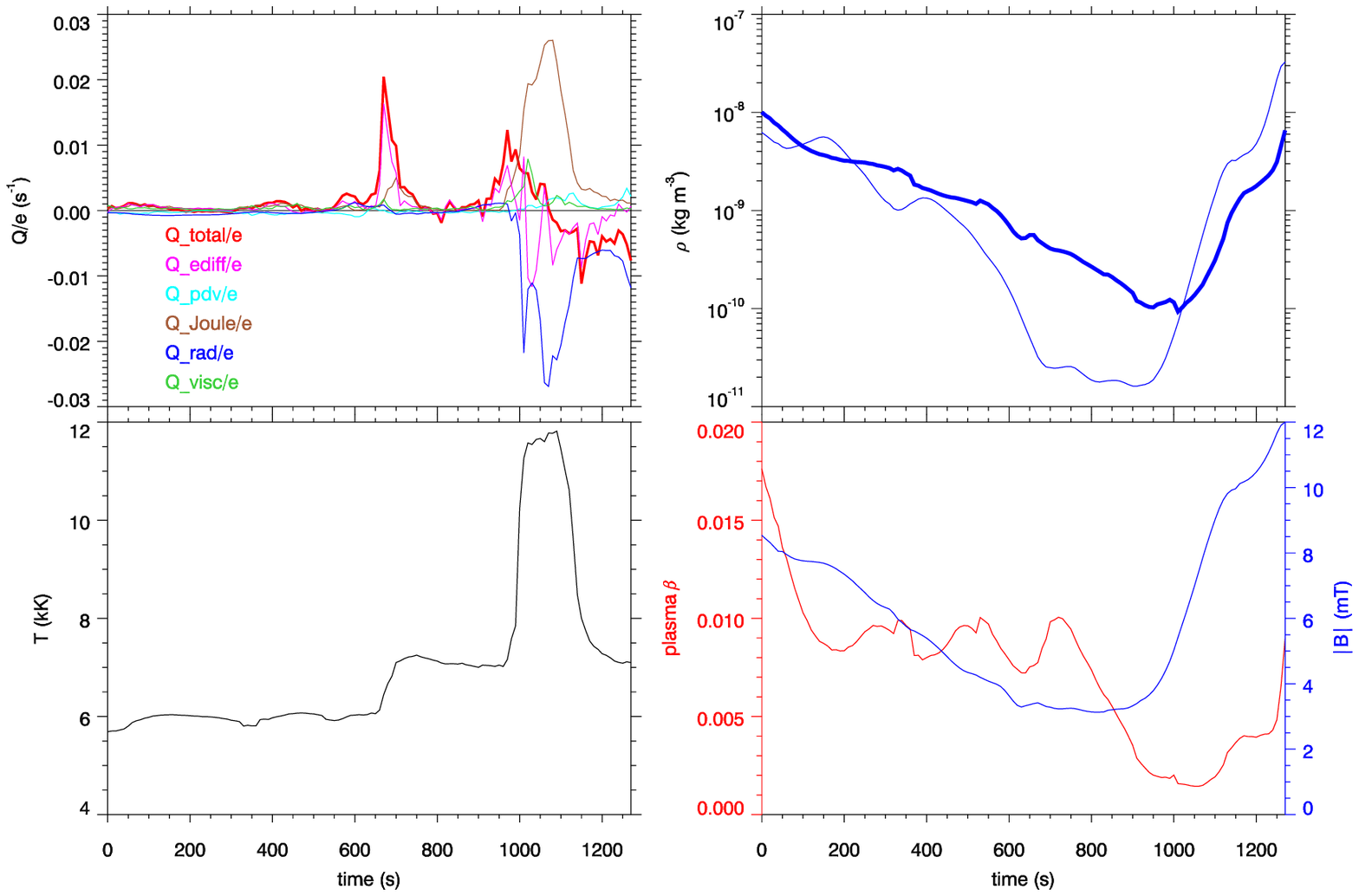}
   \caption{Heating and MHD quantities of the representative cork F1A in Fibril 1 as functions of time.
   {\it Panel a:} heating rate divided by internal energy density $Q/e$ for work done by energy diffusion (magenta), compression (cyan), Joule heating (brown), radiation (blue), viscosity (green), and their sum (red).
   {\it Panel b:} mass density (thick blue), as well as the average mass density at the height of the cork (thin blue).
    {\it Panel c:} temperature.
    {\it Panel d:} magnetic field strength (blue) and plasma~$\beta$ (red).}
   \label{fig:fib1a_evol}
   \end{figure*}



We discuss the behaviour of cork F1A, which drained to a fibril footpoint, in Figs.~\ref{fig:fib1a_acc}\,--\,\ref{fig:fib1a_evol}.

Before the fibril formed, the cork was situated in the low chromosphere and slowly rose from $z=1.2$~Mm to $z=1.8$~Mm during the first 600~s (Fig.~\ref{fig:fib1a_acc}a). During this time the density dropped by an order of magnitude but stays relatively in line with the surrounding plasma at the same height (Fig.~\ref{fig:fib1a_evol}b). The temperature of the material stays relatively constant (Fig.~\ref{fig:fib1a_evol}c). At $t=600$~s a strong p-mode oscillation passes upward. The cork received a strong upwards net Lorenz force (Fig.~\ref{fig:fib1a_acc}e and~\ref{fig:fib1a_acc}h) and rose to around $z=2.7$~Mm over the next 100~s with a peak vertical velocity of 10~km~s$^{-1}$. As confirmed by Fig.~\ref{fig:fib1_spatial}, the Lorentz force is acting roughly perpendicular to the local fieldline. The vertical gas pressure gradient reduced during this period, and the sum of gravity, gas pressure, and Lorentz forces accounted well for the total acceleration. The drop in gas pressure gradient towards zero caused the gravitational deceleration of the material to its quasi "rest" position in the fibril between $t=650$~s and $t=700$~s (Fig.~\ref{fig:fib1a_acc}h).

A decomposition of the Lorentz force into magnetic tension and pressure (not shown in this paper) revealed that the cork experiences an upward force due to an increasing upward magnetic pressure that is not quite balanced by a slightly more slowly increasing downward magnetic tension.

The motion tangential to the fieldline was rather consistent and gentle before and during the mass loading (Fig.~\ref{fig:fib1a_acc}c). The top of the fieldline was flattened throughout and thus there was almost zero net tangential force from gravity force over this time (Fig.~\ref{fig:fib1a_acc}f). 

During the loading of the fibril the density of the cork decreases, but by much less than the mean density of the surrounding atmosphere, leading to the typical over-dense fibrilar structure (Fig.~\ref{fig:fib1a_evol}b).
Joule heating and energy diffusion from the surrounding hot plasma increased the temperature of the cork from  6~kK to 7~kK over the course of 50~s starting at $t=650$~s (Fig.~\ref{fig:fib1a_evol}a and~\ref{fig:fib1a_evol}c).

The cork was relatively stable within the fibril from $t=800$~s to $t=1000$~s, rather than falling back down after the peak of the box oscillation excursion (Fig.~\ref{fig:fib1a_acc}a), instead falling at only a few \kms, but, critically, continuing to decrease in density by nearly an order of magnitude. 

The prominent acceleration tangential to the field began just after $t=700$~s (Fig.~\ref{fig:fib1a_acc}i), under the influence of gravity, which overcame the upward action of the gas pressure gradient when the angle of the fieldline became more vertical, as can be inferred from the increase of the size of the gravitational component at this time. Note that the positive direction is defined relative to the tangential orientation of the field vector - in this case the positive direction points approximately to the left.

By $t=1000$~s, the density had dropped to $10^{-10}$kg~m$^{-3}$ which led to runaway heating that destroyed the fibril. We note that a more extensive survey was conducted for this work, and often during the destruction of dark fibrils one could mistake the disappearing "tails" of the fibrils as draining material. There was certainly material draining from these structures, but it was often the runaway heating and expansion of the material that chases the visible "tail" of the fibril downwards, much faster than the actual draining motions of the material.

It is interesting that the tangential Lorentz force obtains a somewhat non-negligible values between around 
$t=1000$~s and $t=1100$~s (Fig.~\ref{fig:fib1a_acc}i). This was surprising as the Lorentz force is usually negligible in the tangential direction to the field line because the term comes from $\mathbf{j\times B}$ and thus is perpendicular to the magnetic field vector. It is observed that in this case and the other instances of non-zero tangential Lorentz force terms, the non-perpendicular components are linked to strong Joule heating events. Non-zero tangential values can come from the finite precision of the calculations, errors associated with this are greater when the magnetic pressure and tension are exceptionally large with respect to their resultant sum, i.e. the Lorentz force. This situation is most acute during the destruction of the fibrils. Other sources of errors come from small inaccuracies in the calculation of the Frenet-Serret vectors, and grid resolution and interpolation.

The first cork we follow was, however, caught in the draining material and thus avoided much of the heating. At $t=1000$~s the material element was heated in a few tens of seconds from 7~kK to 12~kK (Fig.~\ref{fig:fib1a_evol}). The sources of the heating event are the same as in the previous heating that occurred when the fibril was forming, again with Joule heating being the clearly dominant process due to the currents passing through this hotter, less dense material. Radiative losses kept the material from reaching temperatures where hydrogen would be completely ionised, and then cooled the plasma back down to 7~kK from $t=1100$~s to $t=1200$~s. When a fibril was destroyed in this manner (including all subsequent cases) the magnetic pressure upwards and magnetic tension downwards increase sharply in magnitude, but the resultant magnitude of their sum does not alter particularly. This strong increase of the magnetic pressure gradient causes the separation of nearby fieldlines that is linked to the destruction of the fibrilar structures. As the cork drains out of the fibril, note that it becomes under-dense with respect to the average for material at a similar height (Fig.~\ref{fig:fib1a_evol}b, $t>1000$~s).


In the more horizontal of the transverse directions (here the $\vec{N}$-direction, Fig.~\ref{fig:fib1a_acc}d, g, and j), fibril formation is associated with an oscillating behaviour. This occurred as the cork was loaded to the fibrilar heights (between$t=600$~s and $t=800$~s. This oscillation results from the Lorentz force and are thus likely Alfv\'{e}nic waves propagating along the fibril. The oscillations in this case have a period of $\sim80$~s and decay quickly during the first period from a velocity amplitude of 10 km~s$^{-1}$ to 2 km~s$^{-1}$ and then this amplitude perseveres for another period and a half before they are overcome by the increasing gas pressure gradient force at $t=800$~s, just before the draining of the fibril.

\subsubsection{Fibril 1, Cork B: Runaway heating} \label{sec:F1B}

   \begin{figure*}
   \centering
   \includegraphics[width=0.9\textwidth]{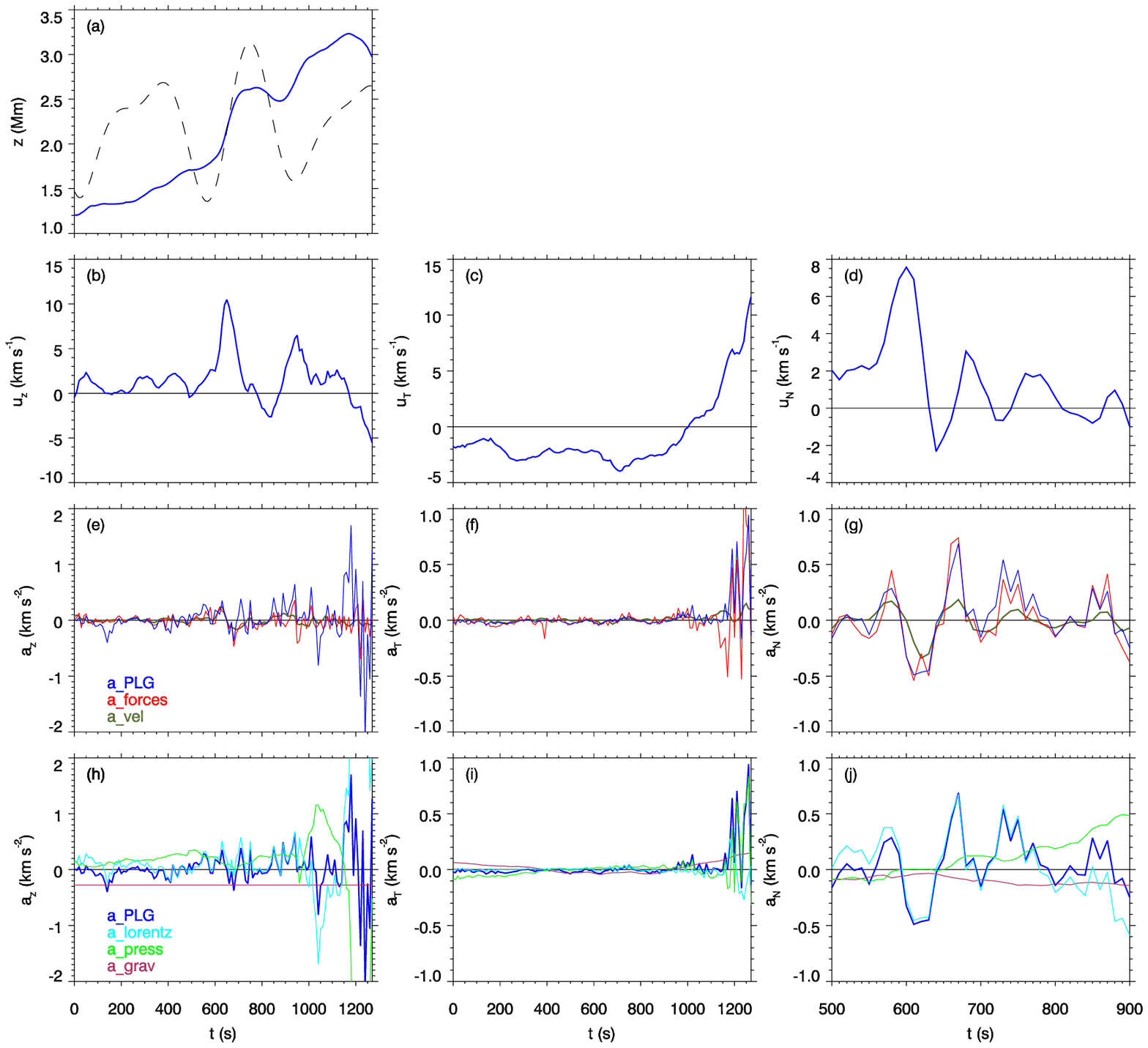}
   \caption{Position, velocity, and acceleration in the $z$-direction of representative cork F1B in Fibril 1.
   Format as in Fig.~\ref{fig:fib1a_acc}.
   }
   \label{fig:fib1b_acc}
   \end{figure*}

   \begin{figure*}
   \centering
   \includegraphics[width=\textwidth]{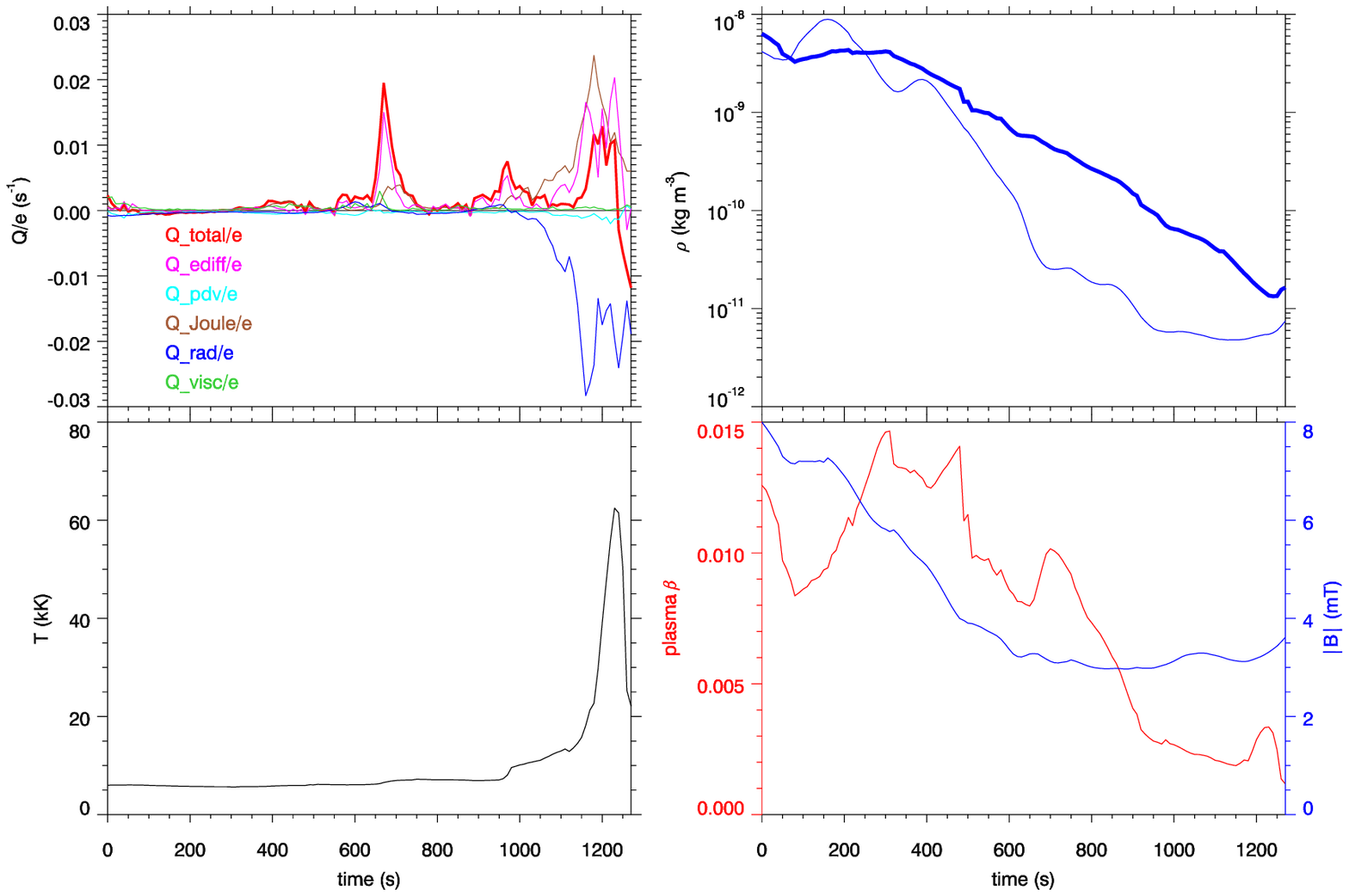}
   \caption{
   Heating and MHD quantities of the representative cork F1B in Fibril 1.
   Format as in Fig.~\ref{fig:fib1a_evol}.
   }
   \label{fig:fib1b_evol}
   \end{figure*}

Next we inspect a cork that does not successfully drain from the fibril but is trapped at the top of the structure when the density drops towards coronal values (Figs.~\ref{fig:fib1b_acc} and \ref{fig:fib1b_evol}).

Like cork F1A, this cork was also not initially located at the footpoint of a loop, but more centrally on the field line loops that rose at around $t=600$~s (Fig.~\ref{fig:fib1_spatial}). The same general values of height increase occurred due to the Lorentz force. The same pattern of oscillations caused by the Lorentz force variations then occurred in the $\vec{N}$-direction.

However, at the location of this cork the fieldline became more vertical after $t=900$~s, so there was no rapid acceleration of the material in a direction tangential to the fieldline at $t=700$~s (Fig.~\ref{fig:fib1b_acc}f). This resulted in the cork not draining down the fieldline as quickly as was the case for cork F1A. Instead, cork F1B continued to rise (Fig.~\ref{fig:fib1b_acc}a) when the next pressure mode hits, and the density of the material decreased (Fig.~\ref{fig:fib1b_evol}b). After $t=900$~s Joule heating increased sharply (Fig.~\ref{fig:fib1b_evol}a, c), which heated, expanded, and ionised the material further. This runaway effect resulted in a sharp spike in temperature shortly after $t=1000$~s when the hydrogen was fully ionised, with the material reaching transition region temperatures of around 60~kK at  $t=1200$~s . The selection of the two corks illustrates the two main mechanisms through which material leaves the fibrils in our experiment: the majority by draining, and a smaller fraction though Joule heating and energy diffusion to transition region temperatures.

To quantify these proportions, we estimated the fraction of mass in the fibril at $t=750$~s that reaches transition region or coronal temperatures at any time during the experiment.
To do so, we assigned a mass to each cork in the same manner as in \citet{2018Zacharias}. That is, by taking the mass density at the seed time ($t=750$~s), multiplying it by the volume of each grid cell, and then equally dividing this mass between each cork present in that grid cell. The total mass present in the fibril was $6.675 \times 10^7$~kg. We then computed the fraction of the total mass in a given temperature range as function of time by adding up the masses assigned to each cork in a given temperature bin.
Table.\ref{tab:fib1} shows the origins and destinations of the mass elements selected in Fibril 1. 

One can see that for this fibril, the mass was fed from the lower chromosphere where the temperature is less than 10~kK. By the end of the experiment most of the fibrilar material had successfully drained down towards chromospheric temperatures, but 12\% is still hotter than 10~kK. 0.036\% is hotter than 50~kK, representing 24\,000~kg. This hot material still appears to be draining and the mass is represented by only a few corks. The fibrils that lift and drain are not very effective suppliers of mass to the transition region and/or corona. 

However during the destruction of the fibril, 3\% of the mass obtained temperatures over 30~kK and 1\% over 50~kK (see Table.~\ref{tab:fib1b}) that could make them temporarily visible in transition region observations. This signature of the heated tail of formerly fibrilar material is a potential observable via simultaneous observations in H$\alpha$ and transition region lines. This 3\% mass fraction would occupy a cube with side length 600~km assuming a lower corona/transition region mass density of $10^{-11}$~kg~m$^{-3}$, which would make it visible in SDO 304 observations.


The draining of fibrilar material leads to lower density loops segments that undergo Joule heating, which provides a possible explanation for transient brightenings of low lying loop sections in TR lines without strong associated footpoint heating, as well as a lack of connectivity to the corona in these events \citep{2014Hansteen_UFS, 2018Pereira_UFS}. Our simulations do not link these events with y-shaped jets above them \citep{2018Pereira_UFS}, although in some cases strong upward motions are caused during the destruction of the fibril. We discuss this using Fibril 2.

\begin{table}
\begin{center}
 \begin{tabular}{l r r r} 
 \hline\hline
 T (kK) & $t=0$ & $t=t_\mathrm{seed}$ & $t=t_\mathrm{max}$ \\ 
 \hline
 $<10$  & 99.26\% & 100\% & 88.03\%  \\
 10-20  &  0.56\% &   0\% & 10.17\%  \\ 
 20-30  &  0.02\% &   0\% &  1.40\%  \\
 30-50  &  0.05\% &   0\% &  0.35\%  \\
 50-100 &  0.11\% &   0\% &  0.02\%  \\
 $>100$ & <0.01\% &   0\% &  0.02\%  \\ 
\hline
 \end{tabular}
 \caption{Origins and destinations of the mass in fibril 1. The percentage of the $6.675 \times 10^7$~kg total mass in Fibril 1 with a temperature in the bins shown in the left column is shown at the beginning (2nd column), seed time (3rd), and at end of the experiment (4th).}
 \label{tab:fib1}
 \end{center}
\end{table}

\begin{table}
\begin{center}
 \begin{tabular}{l r r} 
 \hline\hline
 T (kK) & $t < t_\mathrm{seed}$ & $t > t_\mathrm{seed}$  \\ 
 \hline
 $<10$  & 98.28\%  & 46.66\%   \\
 10-20  &  1.52\%  & 41.33\%   \\ 
 20-30  & <0.01\%  &  8.91\%   \\
 30-50  &  0.05\%  &  1.89\%   \\
 50-100 &  0.02\%  &  1.19\%   \\
 $>100$ &  0.13\%  &  0.02\%   \\ 
\hline
 \end{tabular}
 \caption{Transient heating in fibril 1. Table showing the percentage of the total mass in Fibril 1 that obtains a maximum temperature shown in the left column before the seed time (middle column) and after the seed time (right column).}
 \label{tab:fib1b}
 \end{center}
\end{table}

\subsection{Fibril 2} \label{sec:fib4}
   \begin{figure*}
   \centering
 \includegraphics[width=\textwidth]{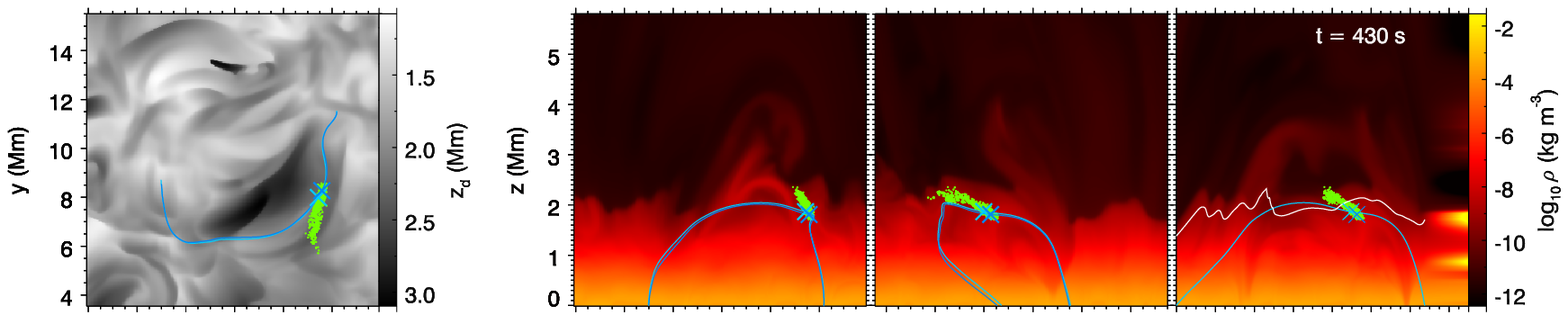}
 \includegraphics[width=\textwidth]{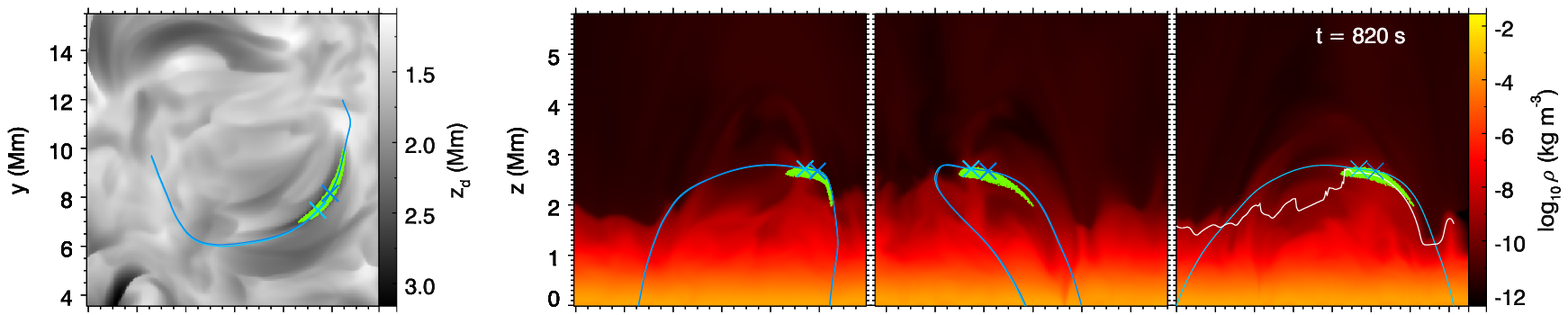}
 \includegraphics[width=\textwidth]{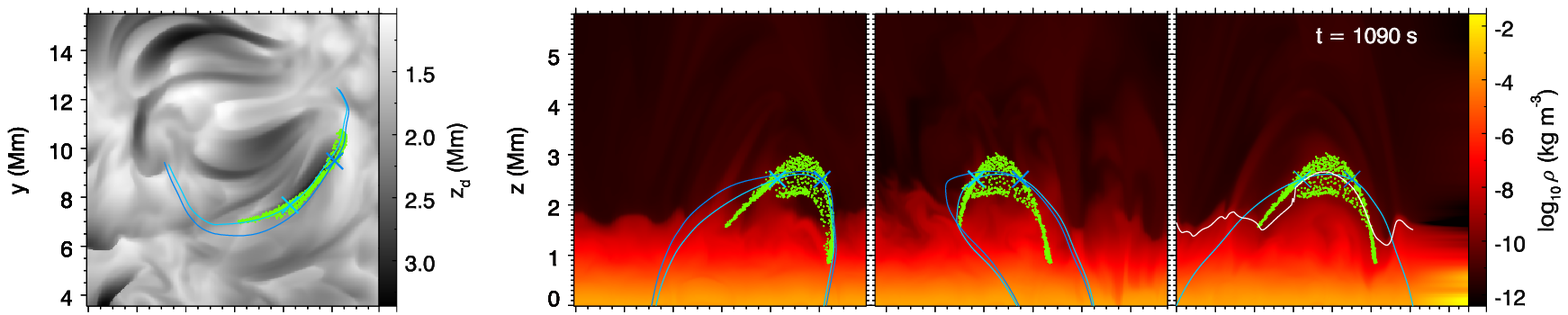}
 \includegraphics[width=\textwidth]{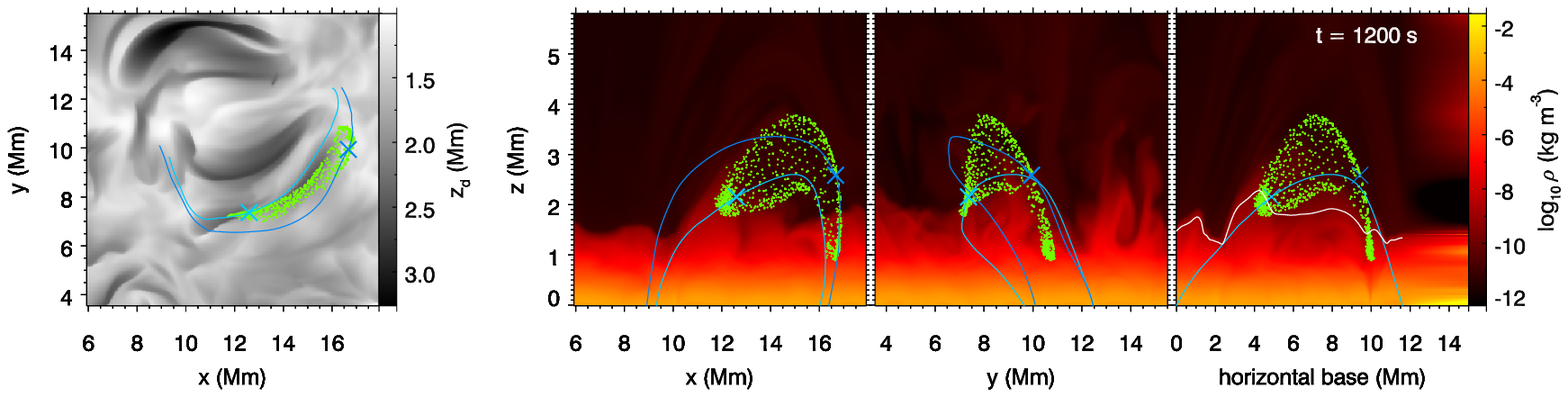}
 \caption{
 Time evolution of Fibril 2. 
 Format as in Fig.~\ref{fig:fib1_spatial}.
 }
   \label{fig:fib2_spatial}
   \end{figure*}

   \begin{figure*}
   \centering
   \includegraphics[width=\textwidth]{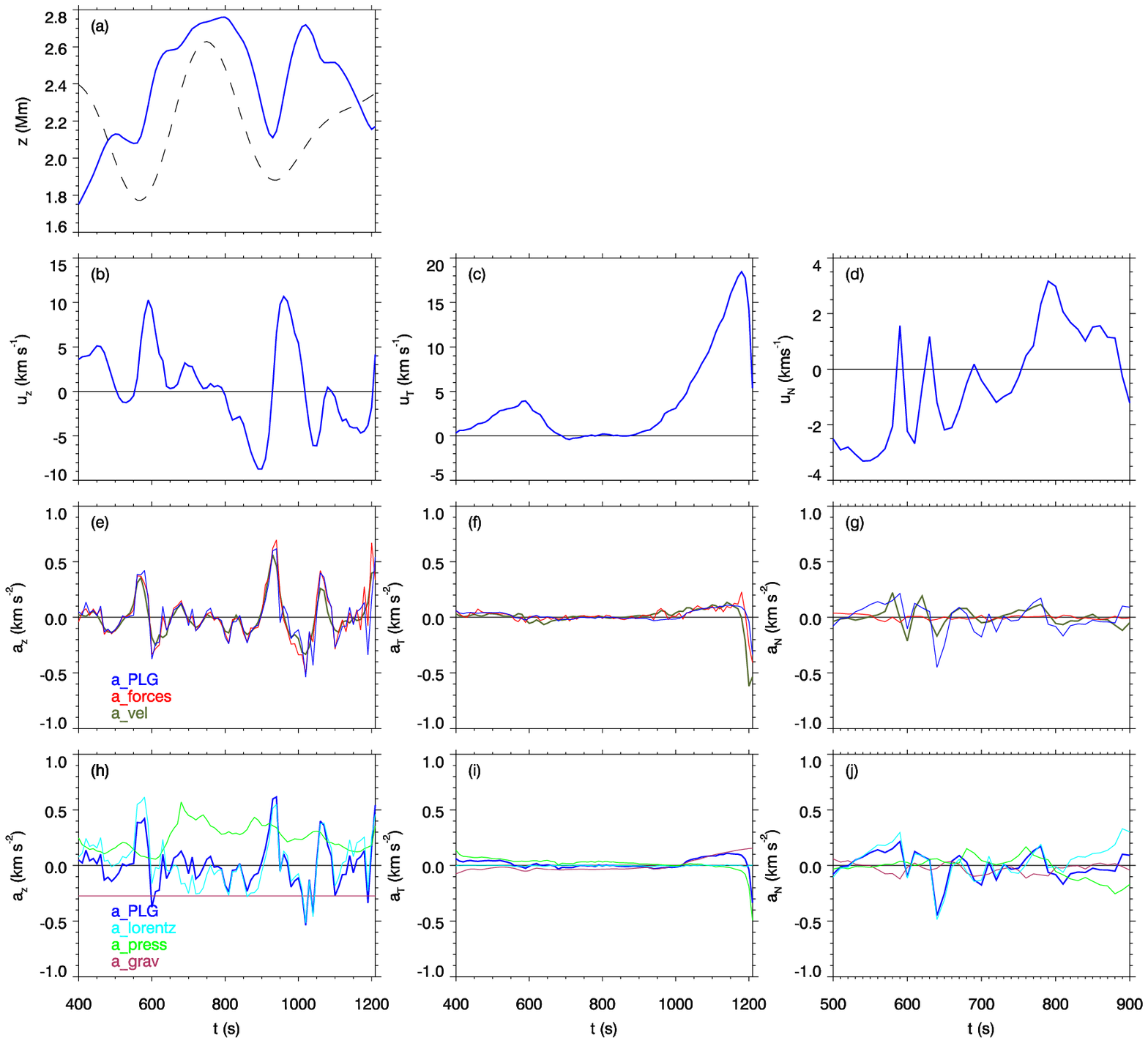}
   \caption{Force component in the z directions for the representative cork A in fibril 2. Format as in Fig~\ref{fig:fib1a_acc}.}
   \label{fig:fib2_acc}
   \end{figure*}


   \begin{figure*}
   \centering
   \subfloat{\includegraphics[width=\textwidth]{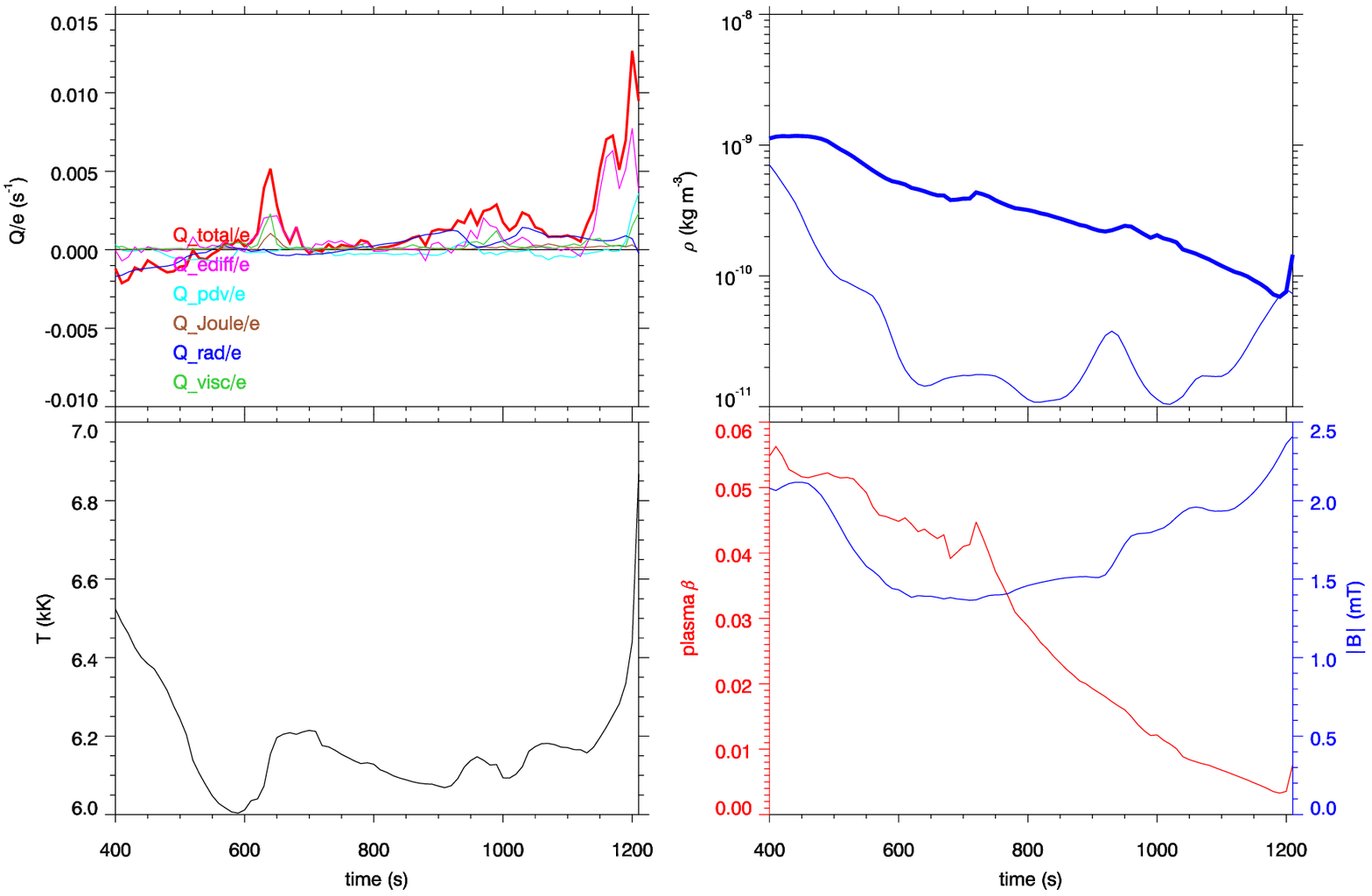}
   }
   \caption{Heating and MHD quantities for the representative cork A in fibril 2. Format as in Fig~\ref{fig:fib1a_evol}.}
   \label{fig:fib2_evol}
   \end{figure*}

The "lift and drain" fibrils are abundant throughout the experiment. However, lift and drain scenarios do not preclude such behaviours as material transiting across the apex of the fieldline, or supplying greater proportions of fibrilar mass to the transition region. An example exhibiting both these behaviours are presented via Fibril~2. For completeness, we present another type of behaviour in Appendix \ref{sec:fib3}: a set of corks that appeared to be loaded up from the footpoint of a static fieldline due to their nearly parabolic motions, but were in fact subject to strong horizontal motions at a similar time to the rising of the fieldline in another lift and drain scenario.


\subsubsection{Fibril 2, Cork A: Traversing the apex of a fibril}

The spatial evolution of Fibril~2 is shown in Fig.~\ref{fig:fib2_spatial}.  The lifting events for this cork in the fibril (at $t=600$~s and $t=950$~s, see Fig.~\ref{fig:fib2_acc}a, b) were again correlated in time with the photospheric box oscillations, although with slightly different offset compared to fibril 1 (the difference in timing is suggestive of the correlation being due to selection bias, although a similar pattern was found for fibrils seeded at $t=500$~s and $t=1000$~s also). In the force plots they were associated with a clear upward Lorentz force as the fieldlines rise and again relax into a more semicircular arc (Fig.~\ref{fig:fib2_acc}e), the key vertical evolution seems largely unrelated to the direct influence of gas pressure forces. 

This was the only event studied where we found that the magnetic pressure gradient and the magnetic tension force both pointed upwards, in contrast to all the other fibrils that we studied (including Fibril~1) where these forces point in opposite directions. We do not show this in the paper, but instead discuss it briefly for completeness.
At $t=600$~s, the magnetic pressure was the larger upward force, but at $t=900$~s it was the magnetic tension. Shortly after the second lifting event, both values diverged rapidly into their more typical values (large magnetic pressure upwards and magnetic tension downwards) as the fibril destruction began.

The motion tangential to the fieldline was unaffected by the Lorentz force (Fig.~\ref{fig:fib2_acc}i). Because a tangential Lorentz force is associated with Joule heating, we expect the temperature of the cork to be largely unaffected by Joule heating. This is precisely what is seen in Fig.~\ref{fig:fib2_evol}, with only one small Joule heating event that did not cause a significant rise in temperature. The cork maintained a relatively constant temperature throughout its journey along the fibril. The formation of dark $H\alpha$ fibrils in our simulations is not intrinsically linked with heating of the material during loading. The motion tangential to the fieldline during loading resulted from an excess of gas pressure over gravity (Fig.~\ref{fig:fib2_acc}i), suggesting that perhaps the p-modes played a direct part in the formation of the fibril, but not noticeably in the $z$-direction. The draining of the cork along the fieldline is again caused by an excess of gravity over the pressure gradient force that began at around $t=900$~s for this cork. The descent of the material is halted by an upward pressure force due to meeting the denser material below (around $t=1200$~s) and the cork by around 500~K.


\begin{table}
\begin{center}
 \begin{tabular}{l r r r} 
 \hline\hline
 T (kK) & $t=0$ & $t=t_\mathrm{seed}$ & $t=t_\mathrm{max}$ \\
 \hline
 $<10$    & 95.5\% & 100\% & 59.3\% \\
 10-20    &  1.6\% &   0\% & 21.9\% \\ 
 20-30    &  0.6\% &   0\% &  3.1\% \\
 30-50    &  0.1\% &   0\% &  1.7\% \\
 50-100   &  0.4\% &   0\% &  2.5\% \\
 100-1000 &  1.8\% &   0\% & 10.8\% \\ 
 >1000    &    0\% &   0\% &  0.7\% \\ 
 \hline
 \end{tabular}
 \caption{ Table showing the origins and destinations of the $8.456 \times 10^7$~kg total mass in fibril 2 with format as in Table \ref{tab:fib1}.}
 \label{tab:fib2}
 \end{center}
\end{table}

\begin{table}
\begin{center}
 \begin{tabular}{l r r} 
 \hline\hline
 T (kK) & $t < t_\mathrm{seed}$ & $t > t_\mathrm{seed}$  \\ 
 \hline
 $<10$    & 94.7\% & 54.0\% \\
 10-20    &  2.1\% & 26.5\% \\ 
 20-30    &  0.9\% &  3.2\% \\
 30-50    &  0.1\% &  1.8\% \\
 50-100   &  0.4\% &  3.0\% \\
 100-1000 &  1.8\% & 10.8\% \\ 
 >1000    &    0\% &  0.7\% \\ 
 \hline
 \end{tabular}
 \caption{ Table showing the extent of transient heating in fibril 2, with format as in Table \ref{tab:fib1b}. }
 \label{tab:fib2b}
 \end{center}
\end{table}

\begin{figure*}
   \centering
   \includegraphics[width=\textwidth]{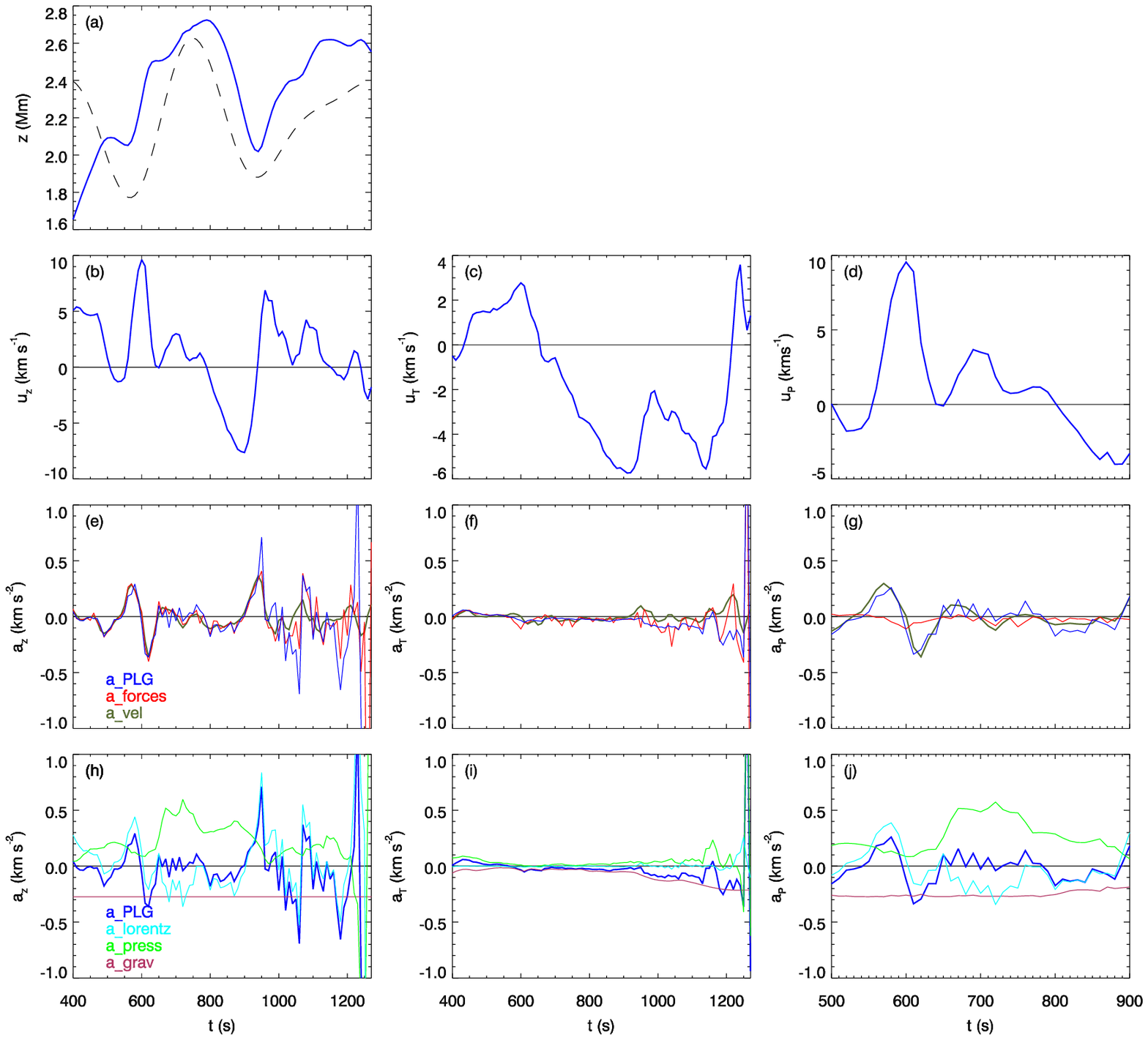}
   \caption{Force component in the z directions for the representative cork B in fibril 2. Format as in Fig~\ref{fig:fib1a_acc}.}
   \label{fig:fib2b_acc}
   \end{figure*}
   \begin{figure*}
   \centering
   \subfloat{\includegraphics[width=\textwidth]{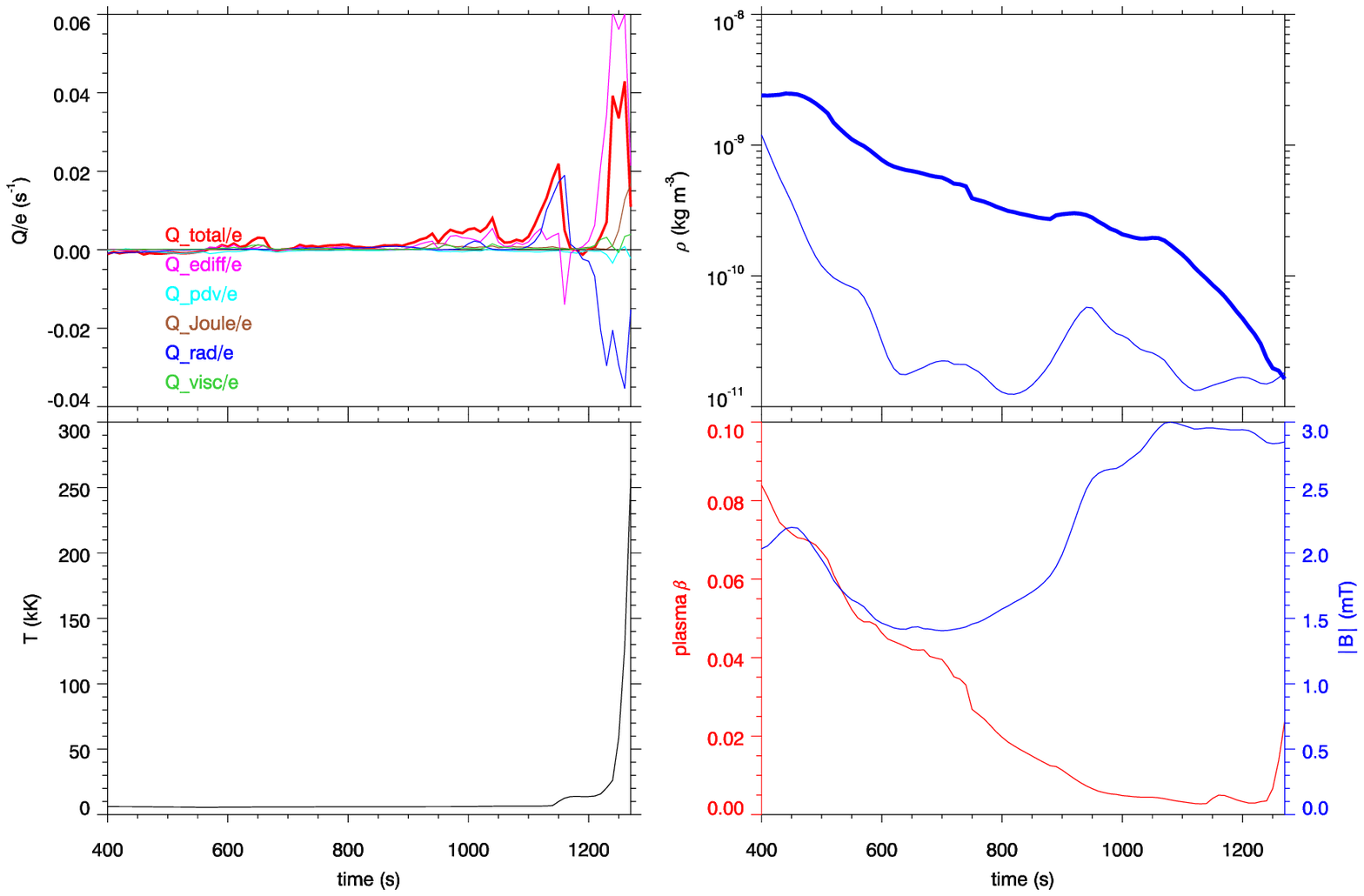}
   }
   \caption{Heating and MHD quantities for the representative cork B in fibril 2. Format as in Fig~\ref{fig:fib1a_evol}.}
   \label{fig:fib2b_evol}
   \end{figure*}

\subsubsection{Fibril 2, Cork B: Runaway heating}

During the destruction of the fibril the angle of the field towards the vertical increased, which resulted in material such as cork A achieving near "footpoint to footpoint" trajectories. However, an increasing pressure forces along some smaller section of the fibril were responsible for the raising of some material high into the atmosphere (see online video of Fig.~\ref{fig:fib2_spatial}). An example of this behaviour is shown via cork B (Figs.\ref{fig:fib2b_acc} and \ref{fig:fib2b_evol}). Although this cork did not rise so high as many others it displays the relevant behaviour. The material expanded (Fig.\ref{fig:fib2b_evol}b, $t=1000-1200$~s) and was then heated by incoming radiation (Fig.\ref{fig:fib2b_evol}a, $t=1100$~s). Finally energy diffusion from the surrounding hot plasma and Joule heating caused a dramatic temperature increase up towards coronal values at the very end of the experiment (Fig.\ref{fig:fib2b_evol}a and c, after $t=1200$~s).

This action of the gas pressure gradient during the destruction of the fibril meant that a much higher proportion of mass was heated to transition region temperatures than for the other simulated fibrils. Tables \ref{tab:fib2} and \ref{tab:fib2b} illustrate this fact. 

Around 97\% of the material is supplied from chromospheric sources, around 95\% coming from the lower chromosphere, and this figure jumps to over 99\% lower chromospheric origin if you consider the start of the experiment to be around $t=300$~s, so the feeding of the mass into the fibril is similar - almost exclusively from the lower chromosphere. However, in the destruction of the fibril, around 15\% of the $8.456 \times 10^7$~kg of mass obtains and maintains transition region temperatures and sustains them until the end of the experiment, with some half a percent obtaining coronal temperatures. Therefore, in cases such as Fibril 2, where the fibril destruction happens during an upward pressure mode, it was found that the disruption to draining can be an effective supplier of mass to the transition region (Fig.~\ref{fig:fib2_spatial}, and Tables \ref{tab:fib2} and \ref{tab:fib2b}). Assuming similar densities for the transition region and low corona as for Fibril 1, the destruction of this fibril supplies mass that fills a volume the size of a cube with a side length of 1~Mm.

\subsection{Mass loading and fieldline twist}

\begin{figure*}
\centering
\subfloat{\includegraphics[width=\textwidth]{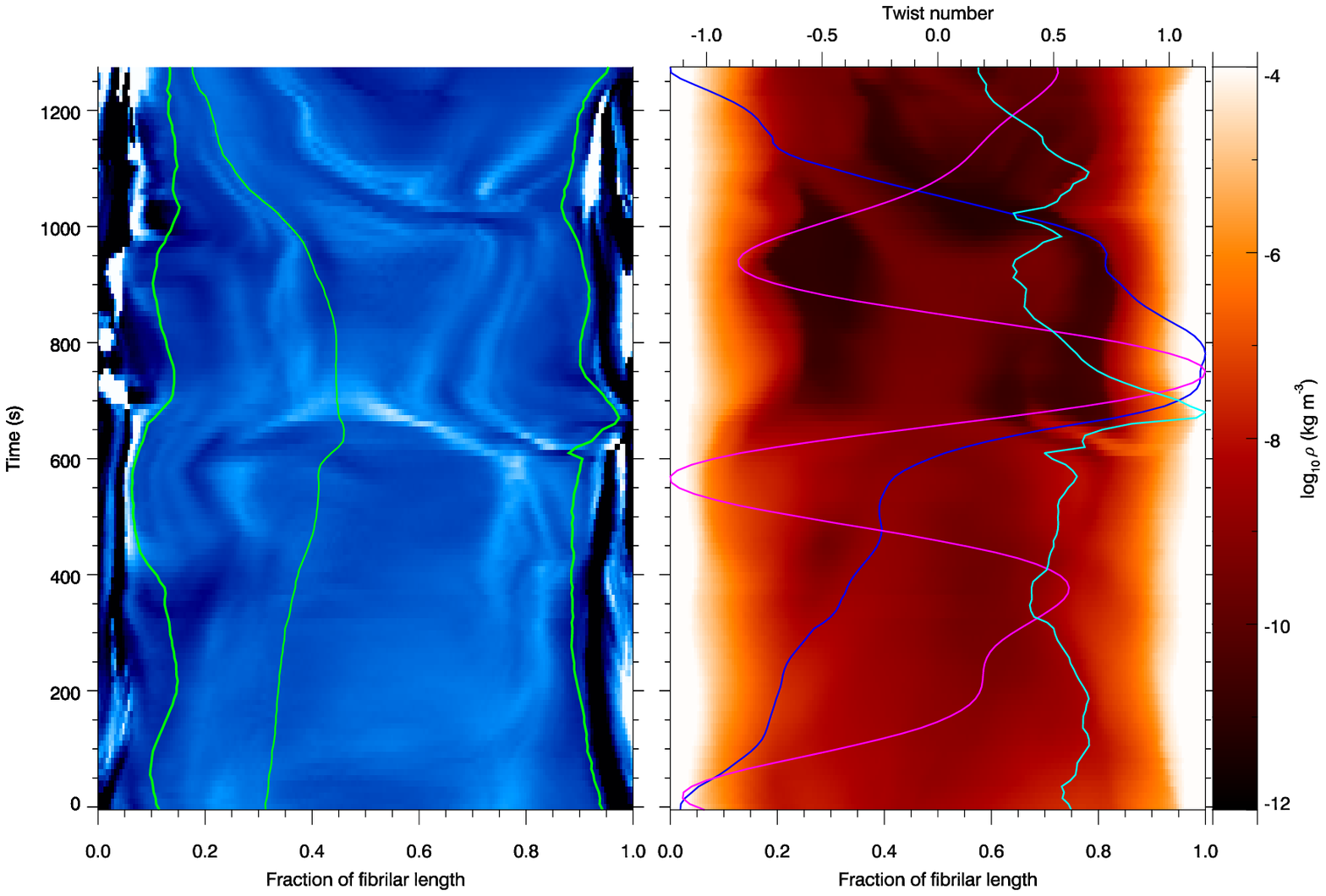}
}
\caption{Fibril formation map for fibril 1. \textit{Left}: The background image shows the twist parameter values along the fieldline length (x-axis. White shows positive values, black negative, and blue shows zero) for the fieldline that passed through cork F1A. The location of the cork along the fieldline is indicated by the central green line. The outer green lines show the edges of the low $\beta$ region, outside of which the twist parameter values are typically larger due to pressure dominating the motions (see the bands of dark and light outside of these green lines). \textit{Right}: The background image shows the logarithm of the plasma density, with overplotted coloured lines that are not scaled similarly to the background image and each other in the x-direction. The magenta line shows the p-mode driven oscillations of the photosphere from Fig.\ref{fig:box_osc}, the blue line shows the vertical z-position of the cork chosen (see Fig.\ref{fig:fib1a_acc}a). The cyan line shows the total integrated twist number along the fieldline, with its axis scaling shown at the top of the image.}
\label{fig:fib3alpha}
\end{figure*}

To look more formally at the untwisting of the fieldlines observed in the previous examples we examine the $\alpha$ parameter, which describes the twist of a fieldline and is defined by the relationship
\begin{equation}
    \mathbf{\nabla} \times \mathbf{ B} = \alpha \mathbf{B}.
\end{equation}
From this relationship the twist parameter can be calculated,
\begin{equation}
    \alpha = \frac{ \left( \mathbf{\nabla} \times \mathbf{B} \right) \cdot \mathbf{B} }{ \lvert \mathbf{B} \rvert ^2 },
\end{equation}
as presented in \cite{2020LinTwist}. We calculate the total twist number of the fieldline for the magnetic twist by integrating the values of this parameter along the fieldline as presented in \cite{2006BergerTwist} in their section 2.4.4 iii. This integral is performed over the section of the magnetic fieldline that passes through the low $\beta$ regime, 
\begin{equation}
    T = \frac{ 1 }{ 4 \pi } \int \alpha \textrm{d}l ,
\end{equation}
where $\textrm{d}l$ is the infinitesimal length along the fieldline at a given point.

We provide a twist parameter map for cork F1A (Fig.\ref{fig:fib3alpha}). Two further maps are provided in the appendices, for cork F2B (Fig.\ref{fig:fib4alpha}) and an example for a fibril that formed earlier in the experiment (Fig.\ref{fig:fibmatsalpha}). In Fig.\ref{fig:fib3alpha} we see again that the main raising event for the material in fibril 1 began just before $t=600$~s (the blue overplotted line in the right panel shows un-scaled vertical position of cork F1A from Fig.\ref{fig:fib1a_acc}a). The fieldlines became destabilised at the same time as a p-mode (magenta line) swept the photosphere upwards. This or some other agent resulted in pulses seen as the tracks in the twist parameter that propagated upwards from near the bases of both footpoints after $t=600$~s (white values in the colour-maps), and a spike in the total twist number for the fieldline (cyan line). The change in twist was also visible in evolution of the field displayed in Fig.\ref{fig:fib1_spatial}, and was accompanied by the beginning of transverse Alfv\'{e}nic wave oscillations of the material, i.e. two or more periods of transverse oscillation with Lorentz force as a restoring force, not shown in this paper. 

Using this information we now attempt to generalise our findings regarding the mass loading of the fibrils in our experiment. 
\begin{enumerate}
    \item Mass is loaded up into the fibrils from the low chromosphere, primarily under the direct action of the Lorentz force. This mass mainly originates from above the apexes of flattened or twisted field loops in the low chromosphere.
    \item These fieldlines were destabilised by actions that changed their local environment. Plausible actions that could produce such changes include alterations in the neighbouring field and granular buffeting. In instances such as fibrils 1 and 2 a plausible destabilising influence was the box oscillations, i.e. p-modes. 
    \item This supposition is based on the temporal correlation between the upswing of the box oscillation between $t=600$~s and $t=750$~s, with the rise in the vertical position of the cork, the visual untwisting of the fieldline, the pulses in alpha twist parameter sent upwards from the edges of the low beta regime at these times, and the evolution of the total twist number for the fieldline. However, this suggestion is also based on the plausible causal relationship between the motions of the plasma near the footpoints of the fieldline and the resulting alterations in the force balance for material on the fieldline that such a motion will cause. Nevertheless we point out that we could not determine a formal deductive link for this supposition and that many actions could cause the fieldlines to destablise. 
    \item Whatever action or actions are the triggers in each individual case, the material rises primarily under the influence of Lorentz force to form high density ridges high in the chromosphere as the field relaxes into more semicircular shapes.
\end{enumerate}

\subsection{Material outside the fibrils} \label{sec:fibno}

\begin{figure}
   \centering
   \includegraphics[width=0.45\textwidth]{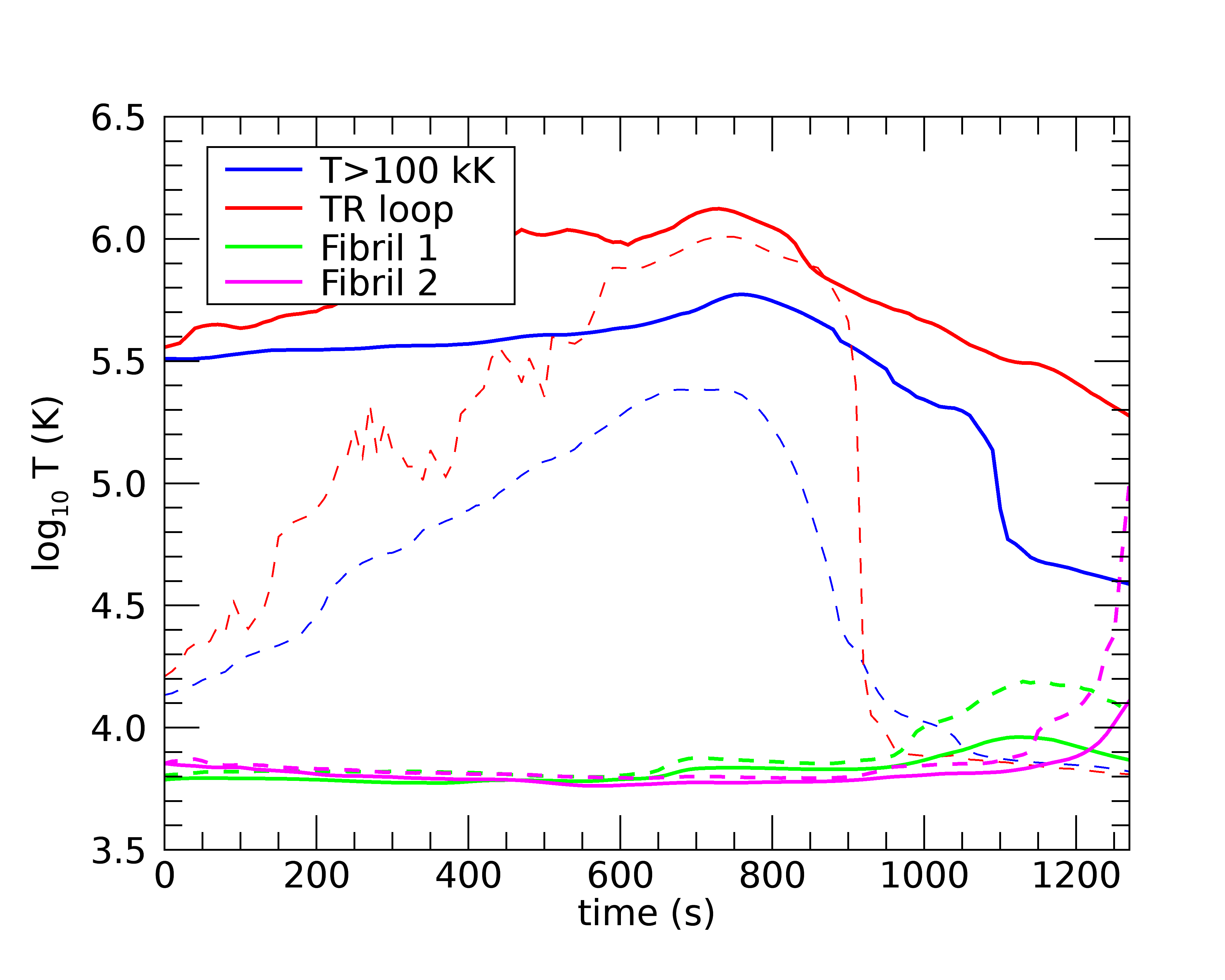}
   \caption{Mean temperature of sets of corks as functions of time. The blue line shows results for a random sample of 100\,000 corks at similar heights as the corks in Fibril 1 \& 2 and which had a temperature above 100K at the seed time. The red line shows the results for corks near an individual fieldline with similar height of the apex to fibrils 1 and 2 but in a low density region. The associated dashed lines show the 10th percentiles, to illustrate cooling due to draining of material towards the chromosphere. Values for the means of the corks in the fibrils are also shown, and their dashed lines indicate the 90th percentile to illustrate heating to transition region temperatures.}
   \label{fig:fibcompare1}
   \end{figure}

In our simulations the dark H$\alpha$ fibrils are high density ridges or bridges of chromospheric material (e.g. Fig.~\ref{fig:fib1_spatial}, third row, right panel). To complete our picture of fibrils it is important to briefly contrast this fibrilar material with the material outside of them. Two examples are presented. Firstly, a statistical sample of 100\,000 corks with heights similar to those at the apexes of Fibrils 1 \& 2 at $t=750$~s and with a temperature above 100~kK. Secondly, the corks within $100~$km of a particular fieldline in a region of low density with a similar apex height to Fibrils 1 \& 2 at $t=750$~s, with only the corks above the minimum height of those corks in Fibril 1 selected. Fig.~\ref{fig:fibcompare1} presents the mean $log_{10}$ temperature evolution of the material in the fibrils alongside the two samples described.

The material in the low density loop stays at relatively high temperatures throughout the experiment. At $t=750$~s the material in the low density loop was at around 1~MK, with the general material outside the fibrils around 500~kK. 

Following these corks backwards in time we see that they were generally at lower temperatures the further back we go, around 300-400~kK at the start of the experiment. The slow increase of the mean temperature between the start of the experiment and the seed time is representative of the corks general drift over time away from the selected hot loop. This brings down the average temperature over time and shows that there is some cycling of material between layers of the atmosphere. At the start of the experiment (over 12 minutes earlier) still, less that 10~\% of the corks were at chromospheric temperatures, seen from the 10th percentiles shown via the dashed lines in Fig.\ref{fig:fibcompare1}.

Moving forward from the seed time, much of the material stays at coronal temperatures but a fraction drains rather more rapidly, dropping the average temperatures of the material down below 100~kK around $t=1100$~s. This draining is somewhat reminiscent of much of the material in the fibrils, albeit at much higher initial temperatures. The statistical sample follows a similar trend to the individual loop, but with somewhat lower average temperatures, indicating that we selected a hotter than average loop by selecting from an area with particularly low density. It is also noticeable that there was a larger proportion of the mass draining from the general hot atmosphere than from the low density loop after around $t=1100$~s.

Comparing this to material at similar heights, but contained within the fibrils we see a very different temperature evolution, where the peak temperature occurs much later in the experiment, and resulted from energy diffusion and Joule heating during the destruction of the fibrils. It is possible that material being heated at the tops of "spicule" structures are the main supplies of material from the chromospheric regions to the transition region and corona. A subsequent investigation will check the flow of mass in more vertical fibrilar structures called mottles. However, material in a fibril that is being destroyed can supply a significant amount of mass if their draining is disrupted by an action such as experiencing an upward pressure mode. This is illustrated via the 90th percentile dotted line for Fibril 2 in Fig.\ref{fig:fibcompare1} at the end of the experiment.

\section{Conclusions} \label{sec:conclusions}

In this study we have used Lagrangian tracer particles to study the mass loading and unloading of fibrils in order to investigate several questions:

\textit{What source(s) supply fibrils with their mass?} 
The origin of the material loaded into the fibrils is almost entirely from the low chromosphere, with a very small proportion coming from the transition region or corona. 

\textit{What are the physical mechanisms/forces that load this mass into the fibrils?} The dominant form of mass loading was a process involving "lift and drain", where the emerging fibrils result from the destabilisation of twisted field lines in the low chromosphere and their subsequent relaxation of field lines raising material trapped over the apexes. Destabilisation of the fieldlines can be triggered by actions including evolution in the neighbouring field, granular buffeting and p-modes. The magnetic pressure gradient increases and the Lorentz force raises the material up through chromosphere to form the elevated over-dense ridges discussed in \cite{2012LeenaartsFibrils}.

This relaxation of the magnetic field allows the fieldlines to gain a greater angle to the horizontal as they become more arched in shape and thus the fibrilar material begins to drain down towards one or both footpoints under gravity. We point out that the motions both of loading and draining studied here refer to actual mass motions, and not to the apparent observational motions of fibrilar structures that can be caused by opacity effects.

It seems that "lift and drain" was also the dominant formation process for fibrils in the simulations of \cite{2012LeenaartsFibrils, 2015LeenaartsFibrils} as evidenced by their Doppler shift diagrams that show blue-shifted material for nearly the entire length of fibrils simultaneously, with red-shifted material draining simultaneously from both ends. This loading process is in contrast to the standard solar picture in which material is given a positive upward velocity from acoustic shocks and guided upwards along relatively static fieldlines from the footpoints. The standard model of mass loading is strongly corroborated by solar observations of fibrils with visible leading edges coming from the bases of the network groups that decelerate over time and can then fall back \citep{2006Hansteen, 2020Kianfar}.

There may well be solar fibrils with mass loaded via the "lift and drain" scenario. Signatures of the scenario would be the observation of simultaneous approximately equal Doppler blue-shifts along the central lengths of a fibril, rather than decreasing Doppler blue-shift of the profiles observed from the footpoint towards the apex of the fibril. This is reminiscent of the rising horizontal field seen in areas of flux emergence, such as the observations of \cite{2002Bernasconi} (see their Fig. 9) with surges and rising centres and draining towards the footpoints \citep{2002Mandrini, 2010Xu}.

It was found that there were some loading scenarios that looked like the standard scenario, but these are also essentially due to the "lift and drain mechanism". The false impression of loading of mass up static fieldlines was generated by strong horizontal or tangential impulses that happen co-temporally with the rising of a fieldline (e.g. Sec.\ref{sec:fib3}). If this process of mass loading is active in the Solar atmosphere it can be identified by the fact that the apparent footpoint will be rooted between magnetic field concentrations, and if the magnetic field vector can be reliably recovered there will also be a strong misalignment between the instantaneous vertical field vector and the motion of the head of the fibril. However, \cite{2015LeenaartsFibrils} also highlight that the vertical magnetic field vector can be misaligned with the fibril orientation for other reasons, e.g. that the fibril forms due to a large group of fieldlines with the apparent body of the fibril tracing out portions of different fieldlines that are in reality offset from each other.

\textit{What processes can destroy fibrils, and what are the dominant forces and energy sources acting on the plasma elements in fibrils?}
In the lift and drain scenario observed in our simulation, the draining phase is closer to that from the classical picture. When the density of the fibril drops towards coronal values, energy diffusion from the hot coronal material surrounding the cork and Joule heating can rapidly increase the temperatures of the remaining material, and thus the visible tail of the fibril in chromospheric lines can "drain" much faster than solar gravity due to this material being heated to transition region temperatures.

\textit{To which locations does the mass that is contained in a fibril flow?} 
Footpoint draining constitutes the overwhelming majority of the destinations of mass in the simulated fibrils. In a typical simulated fibril around 1\% of the mass ever reaches the AIA 304~\AA~sensitivity peak temperature of 50000~K, and this only transiently during the fibril destruction, with less than 0.1\% maintaining this temperature for any length of time. There are exceptions to this, for example when a fibril is struck by a strong pressure mode during its destruction, which can increase the amount of mass supplied to the transition region to over 10\% of the fibrilar mass, and populate large regions of the transition region and corona. 

The fibrils in Bifrost simulations replicate many of the features of solar fibrils such as intensity contrast, Doppler shift magnitudes, and lifetimes \citep{2012LeenaartsFibrils, 2015LeenaartsFibrils}. However, since the dominant mechanism of mass loading in these simulations appears to be different from that from solar observations, it is important to attempt to more closely replicate this. The lack of from-footpoint-static-field loading events could also be the reason that fibrils in Bifrost simulations are so sparsely packed when compared to observations \citep{2015LeenaartsFibrils}. 

What is the cause of this discrepancy, and can it be addressed? 
\begin{enumerate}
    \item Many pressure gradient force events in the positive vertical direction are suppressed by the viscous stress forces in our simulation, so a reduction of the numerical viscosity terms may be important and given the strength of its contribution in this experiment, perhaps energy diffusion should be toned down too.
    \item Cases with more horizontal chromospheric fieldlines showed somewhat more similar evolution to the classic fibril loading mechanism. Therefore perhaps we require more horizontal field structure which could be attained by increasing the domain size, or introducing a global field over the top of the simulation to force  chromospheric field lines trapped beneath to be more horizontal.
    \item Simulation resolution could also play a part if fine structure can help these features to form.
    \item Comparing the previous Bifrost simulations showing fibrils with those from MURaM simulations \citep{2019Bjorgen}, it seems that strong (kG) and complex field structure was always present in regions of the simulation producing densely packed fibrils. It will be interesting to see whether increasing the field strength in our simulations creates more densely packed fibrils.
    \item Perhaps some missing physics such as generalised Ohms law or violations of MHD assumptions such as charge neutrality will play a role in creating fibrils, and modelling that accounts for multi-fluid physics will be necessary.
\end{enumerate}

It is known that mass loading from near the footpoints up relatively static fieldlines occurs on the Sun even from regions with weak magnetic field, so it is important for this to be investigated further in forthcoming simulations to determine the physical mechanisms responsible for this behaviour.

\begin{acknowledgements}
This work was supported by The Swedish Research Council, grant number 2017-04099. JL received support from the Knut and Alice Wallenberg foundation.
The simulations were performed on resources provided by the Swedish National Infrastructure for Computing (SNIC) at the High Performance Computing Center North at Ume\aa~ University, and the PDC Centre for High Performance Computing (PDC-HPC) at the Royal Institute of Technology in Stockholm. This research was supported by the Research Council of Norway through its Centres of Excellence scheme, project number 262622, and through grants of computing time from the Programme for Supercomputing.
\end{acknowledgements}

\bibliographystyle{aa}
\bibliography{corks}

\begin{appendix}
\section{Corks Module details} \label{sec:appcorks}

The corks module \citep{2018LeenaartsCorks, 2018Zacharias} uses passive tracer particles to present the Lagrangian frame, flowing with the mass elements present. In particular the corks module is accurate and versatile because the cork positions are updated on the time-steps of the MHD simulations, rather than using saved data that is stored at intervals much greater than the simulation time-steps \citep{2013ShelyagLagrangeMagneticTornados, 2016NobregaSiverioLagrangeCoolSurge}. The corks pipeline has been parallelised, and updated to enable analysis of the large number of corks that are present in the 3D environment with greatly reduced computing time. The pipeline now also calculates evolution of cork quantities including the forces in the Frenet-Serret coordinate system, and error checks such as searching for duplicate IDs and recording the sizes of jumps that were made in cork coordinates during the pruning and injection processes.

Although the positions and other data for the corks is updated with the hydrodynamic time-steps, this data is only saved to files at the same intervals as the rest of the experiment. A sandbox experiment to check the time-scales on which forces acted upon the corks was conducted. It was confirmed that spikes in the values of particular forces that act on the corks do indeed sometimes occur on time-scales less than 10~s, although the large majority of the acceleration of these tracer particles is described by forces acting on time-scales long enough to be captured in the 10~s snapshots. 

In this experiment the number of corks ($\sim2\times 10^8$) was much greater than in the 2D experiment of \cite{2018LeenaartsCorks} ($\sim4\times 10^5$ corks) in order to avoid unnecessary gluts of corks using up huge amounts of processing and memory, and to avoid voids through which flows could not be traced, the cork injection and pruning processes were switched on, with the snapshot frequency of 10~s. If there was no cork in the cube of grid space closest to any given grid point at the time of the data snapshot, a cork was injected at the grid point. If more than 2 corks were in the same cube, then corks were removed until only two remained. Using 10~s snapshots, around 7\% of the total number corks present are injected at each snapshot.

This benefit for memory and computational cost and the lack of voids in the experiment comes at a price: When constructing pathlines for the corks in time, some fraction of the corks will either have been injected at this point (backward timelines) or pruned (forward timelines). In such a case the pathline jumped to the nearest cork that was not removed in the same snapshot. For forward pathlines (where corks disappear due to being pruned) we are guaranteed that the size of this jump is bounded to $\sqrt{3} \times$ the grid cell separation. A test run produced a mean jump distance of 0.367 grid cells in 3D, with an upper bound almost exactly equal to $\sqrt{3}$ grid cells. Importantly, we wish to discern the sources of the mass loaded onto fibrils, and therefore require accurate backward constructed pathlines. The "removed" corks in these pathlines will have been injected into voids in the experiment, and therefore come with no guarantee that the cork does not jump a large distance. It is therefore important to check that corks are not moving much greater than the distance between grid cells in the time intervals between each sweep of injection and pruning. A test experiment was performed for backward constructed pathlines. The mean jump distance for these corks was 0.563 grid cells, with a positive skewed distribution and some few corks experienced larger jumps. However, only 1 jump from over 46,000 in the test was found to experience a jump over a distance larger than 2 grid cells (2.10 grid cells), and this was in the high corona, well away from the fibrilar features. The majority of the larger jumps come from cells in which a cork has recently (in hydrodynamic time-steps) exited the cell, but none has yet entered it. Therefore, we can be confident of the accuracy of the pathlines in our experiment to a much greater degree than those which are traced using post-processing particles from data that is output to files.

\section{Fibril 3} \label{sec:fib3}

  \begin{figure*}
  \centering
  \includegraphics[width=\textwidth]{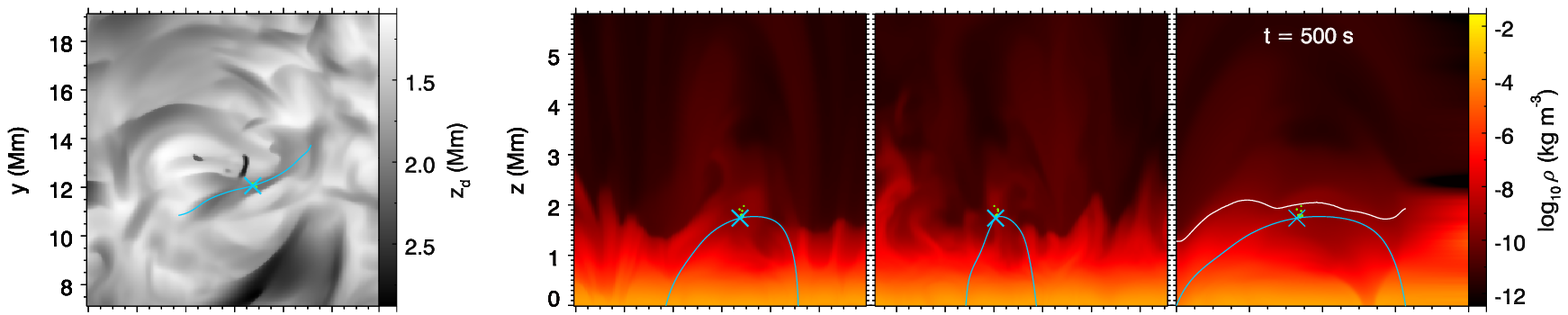}
  \includegraphics[width=\textwidth]{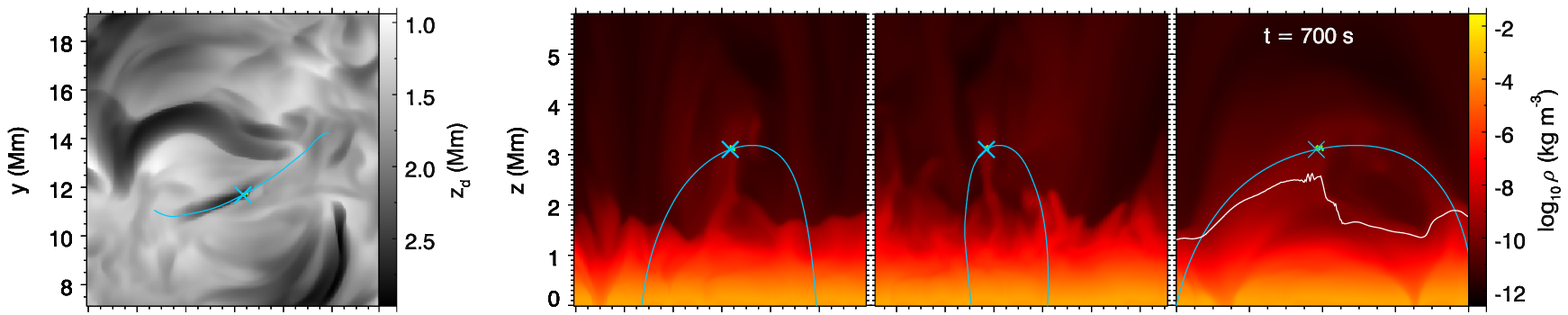}
  \includegraphics[width=\textwidth]{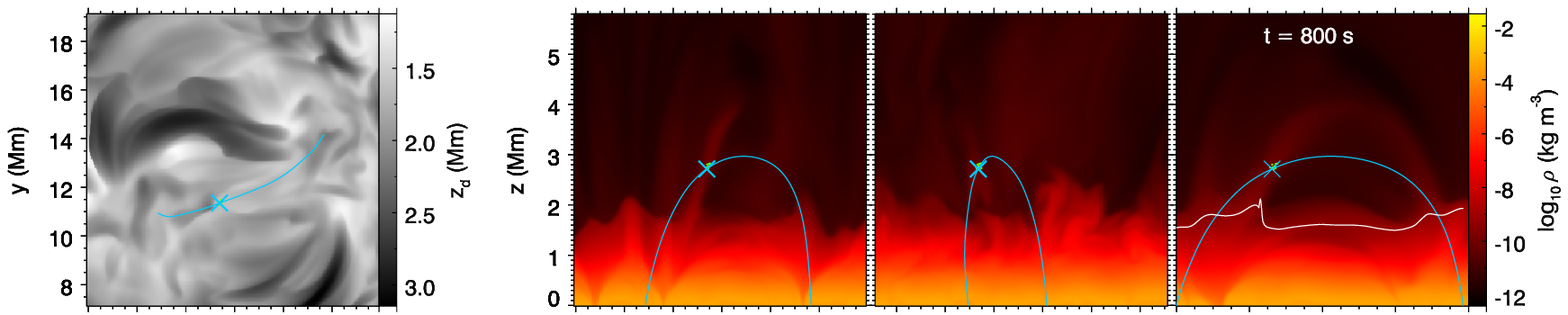}
  \includegraphics[width=\textwidth]{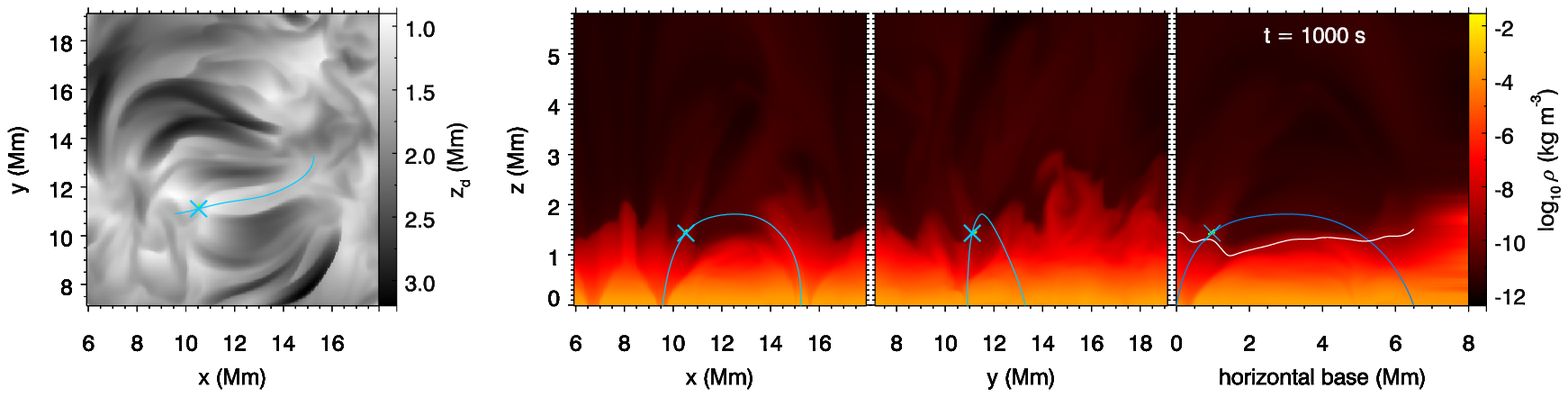}
  \caption{
  Time evolution of Fibril 3. 
  Format as in Fig.~\ref{fig:fib1_spatial}.
  }
  \label{fig:fib3_spatial}
  \end{figure*}

   \begin{figure*}
   \centering
   \includegraphics[width=\textwidth]{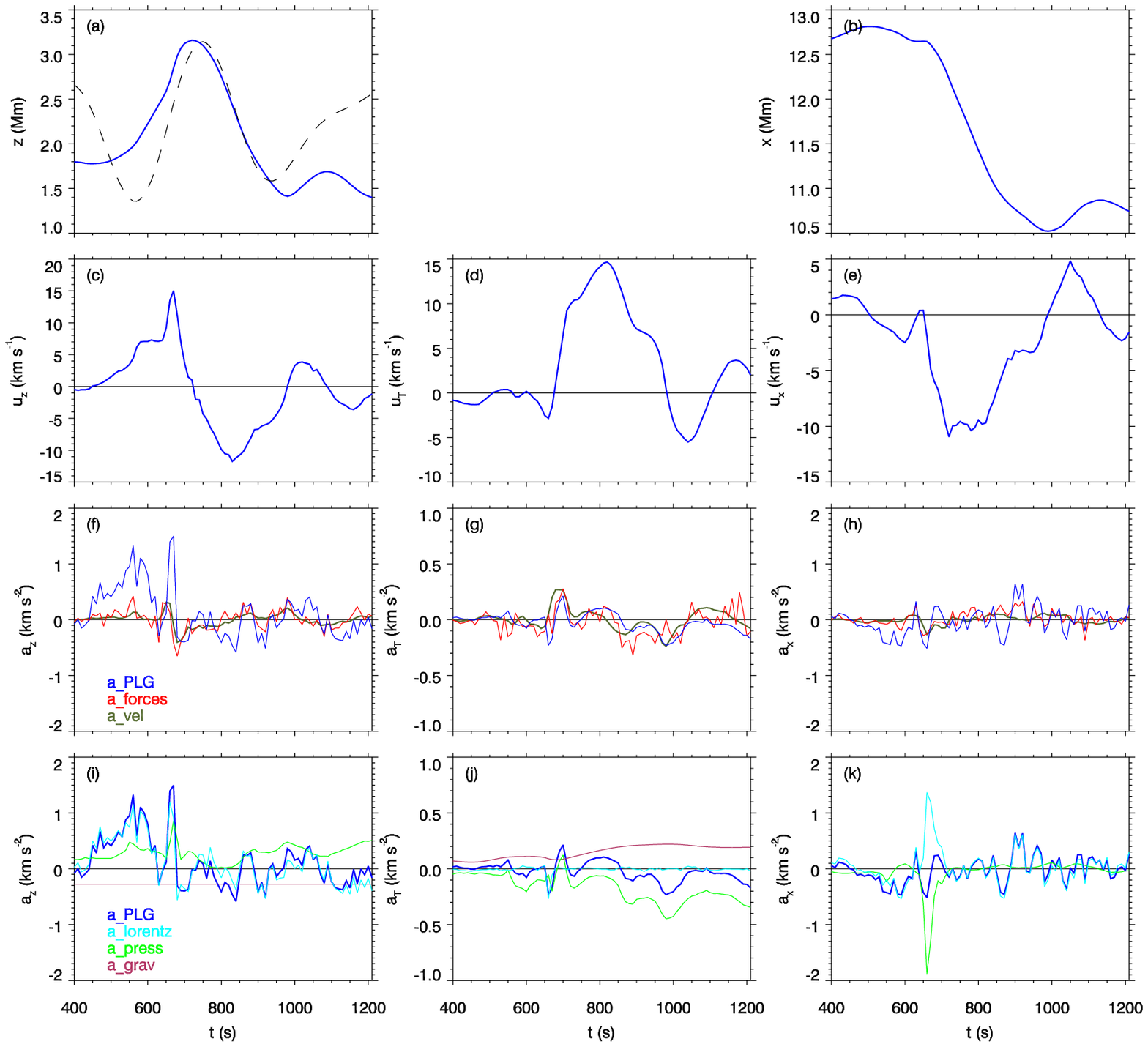}
   \caption{Mean z force components for the 21 corks in fibril 3. Format otherwise as in Fig~\ref{fig:fib1a_acc}.}
   \label{fig:fib3_acc}
   \end{figure*}




There were also fibrilar mass elements showing more "parabolic" loading paths with simultaneous horizontal and vertical motion occurring, as is generally the case in solar observations. One example of which is presented here.

The individual mass elements in fibril 3 underwent a particularly large amount of short term spikes in the Lorentz and viscous stress terms. Therefore we present the averaged results of 21 mass elements which remained adjacent throughout the event, to aid interpretation. 

Fig.~\ref{fig:fib3_spatial} shows the spatial evolution of these 21 mass elements in the fibril, from which we can see that this was still fundamentally a "lift and drain" event, but in this case there was a large horizontal velocity present in the material as the fieldlines rise. There was again significant untwisting of the fieldline that occurs simultaneously with the lifting, this time towards the end of the fibril with lower x values.

Fig.~\ref{fig:fib3_acc} shows the acceleration components in the x-, z-, and magnetic field tangential directions. In this instance the Lorentz and gas pressure gradient forces were working in union to raise the material at around $t=500$~s to $t=600$~s, despite being significantly damped by viscous stress forces. This can be seen by comparing the totals of the Lorentz, gas pressure and gravitational force (blue line, bottom left panel of Fig.~\ref{fig:fib3_acc}f) to the total acceleration from all forces (red/green lines): The upward impulse before $t=600$~s seen in a\_PLG is being significantly countered by the viscous stress force, as can be seen by its inclusion in a\_forces. However, there is some resultant that causes the material to rise, and after $t=600$~s another upwards impulse from the pressure gradient and Lorentz forces raises the material further. Therefore, while the Lorentz force from rising fieldlines is the dominant process of supplying fibrilar mass to the chromosphere in Bifrost simulations, the p-modes can also have a significant direct influence. 

After $t=650$~s a large spike in the gas pressure gradient force acts to accelerate the material in the negative x-direction, against the influence of the Lorentz force (Fig.~\ref{fig:fib3_acc}k). This is confirmed by the noticeable spike in the gas pressure gradient force tangential to the fieldline around the same time (Fig.~\ref{fig:fib3_acc}j. The Lorentz force spike in the x- and tangential directions is again intimately linked to a strong Joule heating event, not presented here.

In this case a mass loading scenario that appeared to be a standard solar scenario, i.e. upward from near the footpoint of a static magnetic fieldline (due to the relatively "parabolic" trajectory), transpired to result from a combination of Lorentz and gas pressure gradient forces, with lateral motions that were dominated by gas pressure gradients, combined with rising fieldlines of the "lift and drain" scenario. This process of mass loading could be distinguished from the general solar scenario of loading in observations with magnetic field analysis because the origin of the material is not at the footpoint of a fieldline rooted in the photosphere. 

\section{Further twist parameter maps}

\begin{figure*}
\centering
\subfloat{\includegraphics[width=\textwidth]{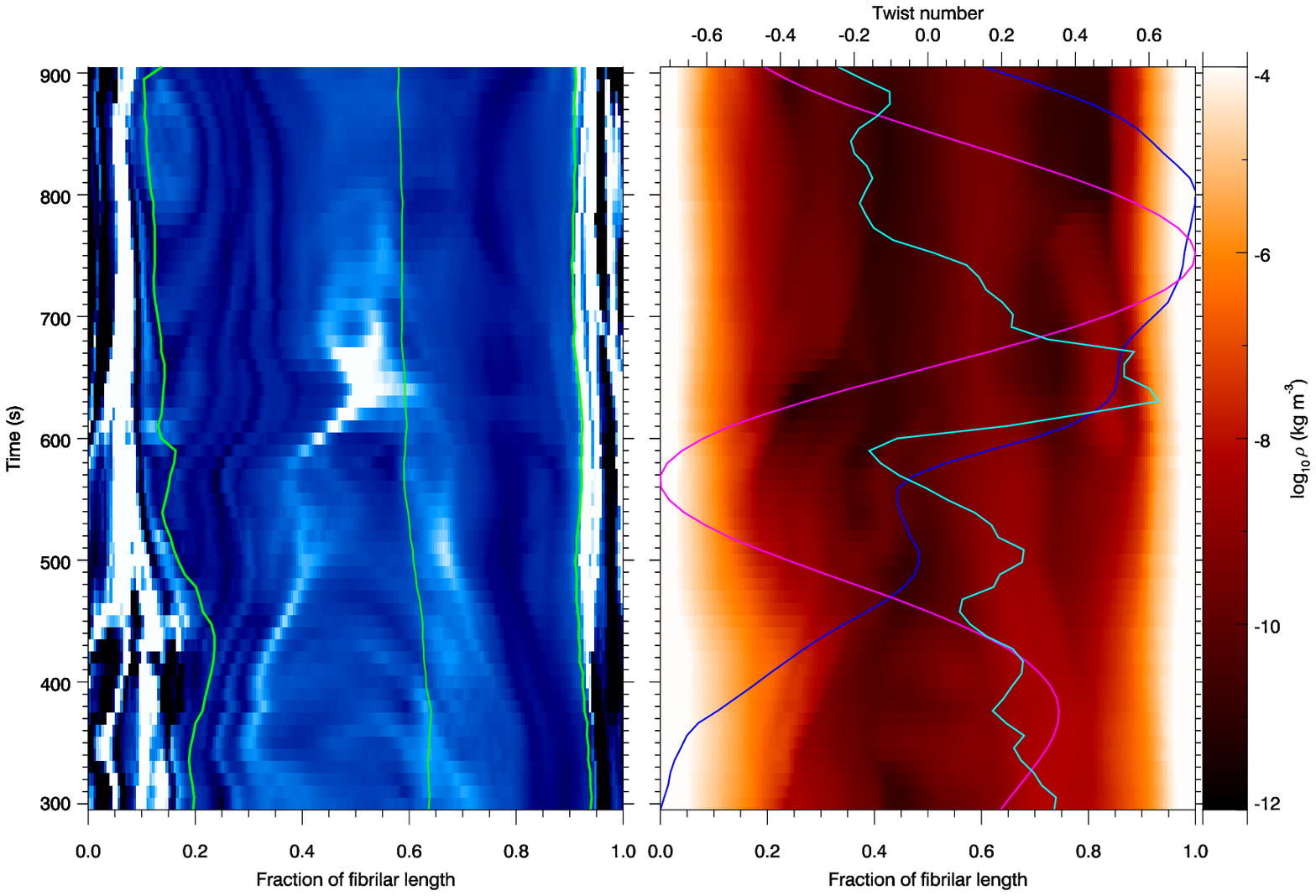}
}
\caption{Fibril formation map for fibril 2. Format as in Fig~\ref{fig:fib3alpha}.}
\label{fig:fib4alpha}
\end{figure*}

The story of the mass loading for fibril 2 was very similar to fibril 1 (Fig.\ref{fig:fib4alpha}), except that the disturbance propagated primarily from one end of the fieldline. 

\begin{figure*}
\centering
\subfloat{\includegraphics[width=\textwidth]{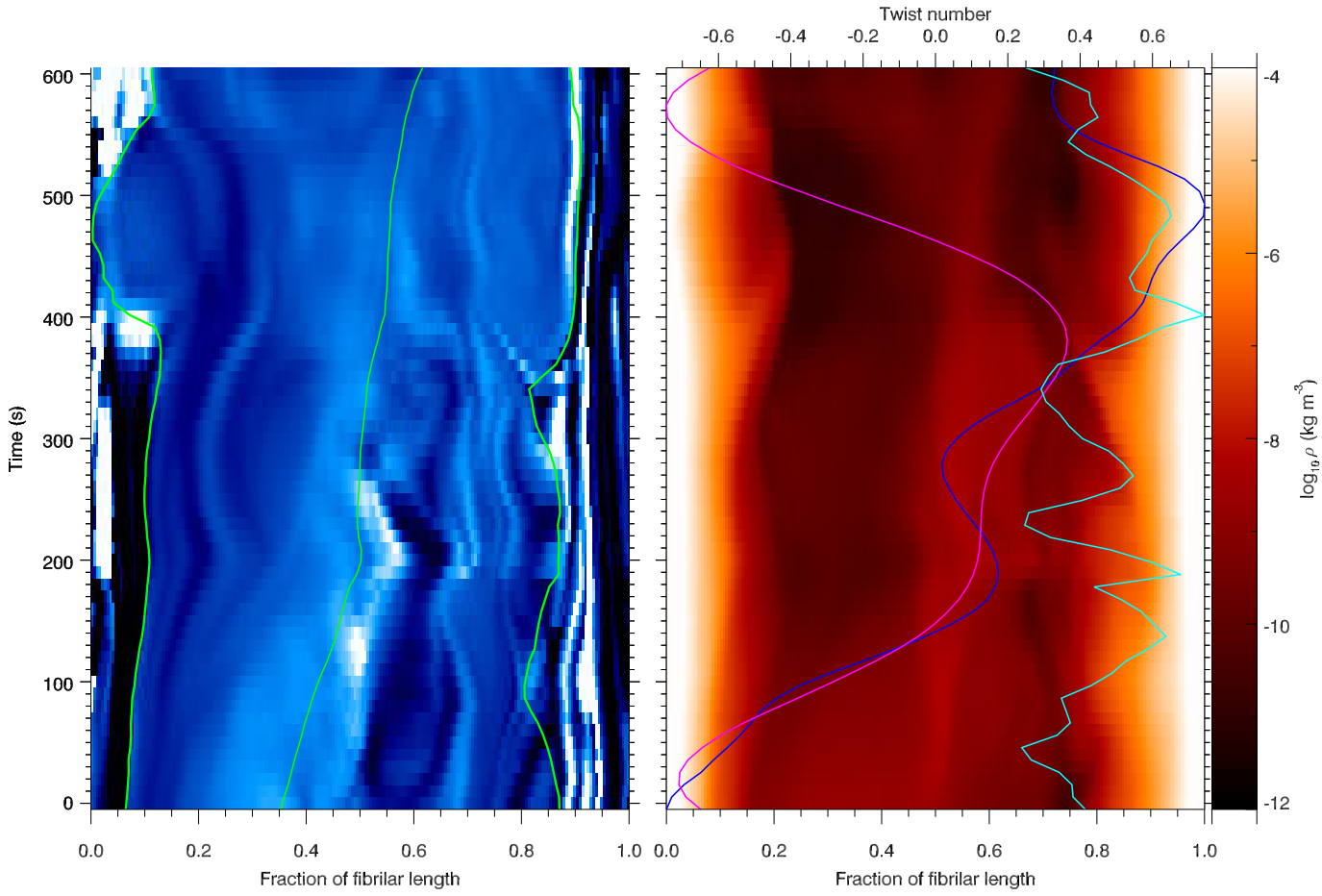}
}
\caption{Fibril formation map for a fibril that formed earlier in the experiment. Format as in Fig~\ref{fig:fib3alpha}.}
\label{fig:fibmatsalpha}
\end{figure*}

The alpha parameter map for a lower lying fibril seeded at $t=500$~s of the experiment is shown in Fig.\ref{fig:fibmatsalpha}. The material in this fibril rose via two weaker actions related to the Lorentz force. The first action, from $t=0$~s to $t=200$~s appears to start slightly before the beginning of the experiment during the model relaxation phase, there are only vague upward pulse tracks on the left side of the alpha parameter plot that correspond with this gradual rise. At the end of this phase there are quite strong positive and negative twists present near the top of the fieldline that persist until around 300s. The second action is fully captured in the time series between $t=300$~s and $t=500$~s. The cause of this action is unclear. It could be linked to the p-modes, whose effects on the base of the low $\beta$ region can be clearly seen towards the left side of the density map (Fig.\ref{fig:fibmatsalpha}, \textit{right}), the cancellation of the opposite twists near the loop apex, or the actions occurring near the boundary of the low $\beta$ region that causes tracks in the density and twist parameter from the right footpoint at around $t=270$~s and from both ends starting at around $t=350$~s.

\end{appendix}

\end{document}